\documentclass{aa}  
\usepackage{graphicx}
\usepackage{txfonts}
\usepackage{natbib}
\bibpunct{(}{)}{;}{a}{}{,} % to follow the A&A style
\begin{document}
  \title{IMAGES\thanks{Intermediate MAss Galaxy Evolution Sequence,
      ESO programs 174.B-0328(A), 174.B-0328(E)}-III: The evolution of
    the Near-Infrared Tully-Fisher relation over the last 6 Gyr}

  \titlerunning{IMAGES-III: evolution of the NIR TFR}

  \author{
  M. Puech\inst{1,2}
  \and H. Flores\inst{2}                   %Hector Flores <hector.flores@obspm.fr>   
  \and F. Hammer\inst{2}                   %Francois Hammer <francois.hammer@obspm.fr>                   
  \and Y. Yang\inst{2}                     %yanbin.yang@obspm.fr
  \and B. Neichel\inst{2}                  %Benoit Neichel <benoit.neichel@obspm.fr>                     
  \and M. Lehnert\inst{2}                  %Matt Lehnert <matthew.lehnert@obspm.fr
  \and L. Chemin\inst{2}                   %laurent.chemin@obspm.fr
  \and N. Nesvadba\inst{2}                 %Nicole Nesvadba <nicole.nesvadba@obspm.fr>                   
  \and B. Epinat\inst{5}
  \and P. Amram\inst{5}                    %Philippe.Amram@oamp.fr   
  \and C. Balkowski\inst{2}                %Chantal Balkowski <Chantal.Balkowski@obspm.fr> 
  \and C. Cesarsky\inst{1}                 %Catherine Cesarsky <ccesarsk@eso.org
  \and H. Dannerbauer\inst{6}              %Helmut Dannerbauer <dannerb@mpia-hd.mpg.de>                  
  \and S. di Serego Alighieri\inst{7}      %Sperello di Serego Alighieri <sperello@arcetri.astro.it>     
  \and I. Fuentes-Carrera\inst{2}          %isaura.fuentes@obspm.fr               
  \and B. Guiderdoni\inst{8}             %Bruno Guiderdoni <bguider@obs.univ-lyon1.fr>
  \and A. Kembhavi\inst{3}                 %Ajit Kembhavi <akk@iucaa.ernet.in>                           
  \and Y. C. Liang\inst{9}                %Yanchun Liang <ycliang@bao.ac.cn>                            
  \and G. {\"O}stlin\inst{10}              %Goeran Ostlin <ostlin@astro.su.se>                           
  \and L. Pozzetti\inst{4}                 %lucia@bo.astro.it                                 
  \and C. D. Ravikumar\inst{11}            %--? cdravi@gmail.com, Chazhiyat.Ravikumar@obspm.fr
  \and A. Rawat\inst{2,3}                  %Abhishek Rawat <rawat@iucaa.ernet.in>
  \and D. Vergani\inst{12}                 %daniela@lambrate.inaf.it
  \and J. Vernet\inst{1}                   %Joel Vernet <jvernet@eso.org>
  \and H. Wozniak\inst{8}                  %herve.wozniak@obs.univ-lyon1.fr
}

   \offprints{mpuech@eso.org}

\institute{
  %1
ESO, Karl-Schwarzschild-Strasse 2, D-85748 Garching bei M\"unchen, Germany
\and  %2
GEPI, Observatoire de Paris, CNRS, University Paris Diderot; 5 Place Jules Janssen, 92190 Meudon, France 
\and %3
Inter-University Centre for Astronomy and Astrophysics, Post Bag 4, Ganeshkhind, Pune 411007, India 
\and %4
 INAF - Osservatorio Astronomico di Bologna, via Ranzani 1, 40127 Bologna, Italy
\and %5
Laboratoire d'Astrophysique de Marseille, Observatoire Astronomique de 
Marseille-Provence, 2 Place Le Verrier, 13248 Marseille, France
\and %6
MPIA, K{\"o}nigstuhl 17, D-69117 Heidelberg, Germany
\and %7
INAF, Osservatorio Astrofisico di Arcetri, Largo Enrico Fermi 5, I-50125, Florence, Italy
\and %8
Centre de Recherche Astronomique de Lyon, 9 Avenue Charles Andr\'e, 69561 Saint-Genis-Laval 
Cedex, France 
\and %9
National Astronomical Observatories, Chinese Academy of Sciences, 20A Datun Road, Chaoyang District, Beijing 100012, PR China 
\and %10
Stockholm Observatory, AlbaNova University Center, Stockholms Center for Physics, Astronomy and Biotechnology, Roslagstullsbacken 21, 10691 Stockholm, Sweden 
\and %11
Department of Physics, University of Calicut, Kerala 673635, India
\and %12
IASF-INAF - via Bassini 15, I-20133, Milano, Italy
}

\date{Received ......; accepted .....}

\abstract{Using the multi-integral field spectrograph GIRAFFE at VLT,
  we have derived the K-band Tully-Fisher relation (TFR) at $z \sim
  0.6$ for a representative sample of 65 galaxies with emission lines
  ($W_\mathrm{0}\mathrm{(OII)}\ge15\AA$). We confirm that the scatter
  in the $z \sim 0.6$ TFR is caused by galaxies with anomalous
  kinematics, and find a positive and strong correlation between the
  complexity of the kinematics and the scatter that they contribute to
  the TFR. Considering only relaxed-rotating disks, the scatter, and
  possibly also the slope, of the TFR, do not appear to evolve with
  redshift. We detect an evolution of the K-band TFR zero point
  between $z \sim 0.6$ and $z=0$, which, if interpreted as an
  evolution of the K-band luminosity of rotating disks, would imply
  that a brightening of $0.66 \pm 0.14$ mag occurs between $z \sim
  0.6$ and $z=0$. Any disagreement with the results of \cite{flores06}
  are attributed to both an improvement of the local TFR and the more
  detailed accurate measurement of the rotation velocities in the
  distant sample. Most of the uncertainty can be explained by the
  relatively coarse spatial-resolution of the kinematical data.
  Because most rotating disks at $z\sim0.6$ are unlikely to experience
  further merging events, one may assume that their rotational
  velocity, which is taken as a proxy of the total mass, does not
  evolve dramatically. If true, our result implies that rotating disks
  observed at $z\sim0.6$ are rapidly transforming their gas into
  stars, to be able to double their stellar masses and be observed on
  the TFR at $z=0$. The rotating disks observed are indeed
  emission-line galaxies that are either starbursts or LIRGs, which
  implies that they are forming stars at a high rate. Thus, a
  significant fraction of the rotating disks are forming the bulk of
  their stars within 6 to 8 Gyr, in good agreement with former studies
  of the evolution of the mass-metallicity relationship.

\keywords{Galaxies: evolution; Galaxies: kinematics and dynamics;
   Galaxies: high-redshifts; galaxies: general; galaxies:
   interactions; galaxies: spiral.}
}

\maketitle

\section{Introduction}
Since the first rotation curves were measured at intermediate
redshifts \citep{vogt93}, many studies have been devoted to the
evolution of the Tully-Fisher Relation (TFR, \citealt{tully77}), given
its prominent role in constraining galaxy-formation models (e.g.,
\citealt{dutton06}). Early work using B-band imaging found only modest
luminosity evolution \citep{vogt96,vogt97}, but subsequent studies
using data of the same band have suggested a more significant
evolution \citep{simard98,bamford06,weiner06,chiu07}. In almost all of
these studies, the B-band TFR shows a large dispersion in comparison
with the local relation, especially at the low-luminosity (or
velocity) end (e.g., \citealt{bohm04}). \cite{bohm07} showed that this
effect could be attributed to an incompleteness in magnitude provided
that the scatter decreased by a factor of at least three between
$z\sim0.5$ and $z=0$, with no evolution of slope or zero point.
Alternatively, they proposed a possible luminosity-dependent
evolution, in which distant low-luminosity galaxies would have lower
mass-to-light ratios [$M/L$] than their local counterparts, while
higher-luminosity galaxies would not undergo strong $M/L$ evolution.
Therefore, the situation for B-band data remains unclear, particularly
concerning the large scatter in measurements found at high redshift.

To minimize observational biases and enable easier comparisons with
models, a progressive interest has been devoted to the K-band and
stellar-mass TFRs (hereafter, smTFR). Long-slit spectroscopy of
distant galaxies revealed an smTFR of significant scatter, with
detectable evolution of neither zero point nor slope, up to $z\sim1$
\citep{conselice05}. This much larger dispersion appears to be linked
to ``kinematically anomalous galaxies'', as inferred from local
studies \citep{kannappan04}, or, from a morphological point of view,
with disturbed, compact, or merging galaxies
\citep{kassin07,atkinson07}. \cite{weiner06} and \cite{kassin07}
defined a new tracer of the galaxy dark-halo potential called $S$,
which corrects the rotation velocity for disordered, non-circular
motions. Once expressed using this new kinematical estimator, the
distant smTFR shows significantly-reduced scatter, with no detectable
evolution in either zero point or slope. This suggests an important
role of non-ordered motions in increasing the scatter of the distant
TFR.

In this respect, 3D spectroscopy provides a unique way to distinguish
relaxed rotating disks from other kinematically-disturbed galaxies.
Kinematically-selected relaxed (or pure) rotating disks at $z\sim0.6$
present a TFR that appears to evolve neither in slope, zero point, nor
\emph{dispersion} \citep{flores06}: kinematically-anomalous galaxies,
which range from perturbed rotators where a rotation can be detected
to galaxies with complex kinematics but no noticeable rotation, appear
to be responsible for the increased scatter of the distant TFR.
Therefore, 3D spectroscopy allows us to establish a direct connection
between non-ordered motions and kinematical and morphological
anomalies. In this paper, we double the sample previously available
(see \citealt{yang07}, hereafter Paper I) to test robustly the results
obtained in \cite{flores06}.

This paper is organized as follows: Sect. 2 introduces the data used
in this study; Sect. 3 details the methodology used for the analysis;
Sect. 4 presents the K-band TFR obtained from GIRAFFE data, and Sect.
5 draw the conclusions from this work. In two appendixes, we derive
the K-band luminosity density and the stellar-mass TFR in the GIRAFFE
sample. Throughout, we adopt $H_0=70$ km/s/Mpc, $\Omega _M=0.3$, and
$\Omega _\Lambda=0.7$, and the $AB$ magnitude system.

\section{Data}

\subsection{Distant sample}

\subsubsection{Kinematics}
We used the multi-object integral field spectrograph FLAMES-GIRAFFE at
VLT, to obtain the [OII] spatially-resolved kinematics of a sample of
68 galaxies at $0.4 \leq z \leq 0.75$. The median redshift of the
sample was $z_{med}\sim0.61$, which corresponds to a look-back time
$\sim5.8$ Gyr, while the 25 and 75 percentiles of the redshift
distribution are $z_{25}\sim0.49$ and $z_{75}\sim0.67$. This sample
represents well the luminosity function of $z\sim0.6$ galaxies with
stellar masses [$M_{stellar}$] in the $1.5-15
\!\times\!10^{10}M_{\sun}$ range, and is unaffected by field-to-field
variations within Poisson statistics, as targets were observed in four
different fields (see Paper I).

\cite{flores06} and Paper I used GIRAFFE velocity fields [VF] and
velocity dispersion maps [$\sigma$-maps], to define three kinematical
classes, namely the \emph{rotating disks} [RD], which correspond to
relaxed rotators, the \emph{perturbed rotators} [PR], corresponding to
rotating disks showing some weak disturbances, and
\emph{kinematically-complex} [CK] galaxies, corresponding to
dynamically non-relaxed galaxies, probably associated with major
mergers (see also \citealt{puech06,puech07}). This classification
takes into account the residuals between the observed VF and
$\sigma$-map and those predicted by a rotating-disk model (see Paper
I), which mitigates the subjectivity of a fully visual classification.

\subsubsection{Morphology}
A detailed morphological analysis of a subsample of 52 galaxies, which
had multi-band HST/ACS imaging data, was completed by \cite{neichel07}
(hereafter, Paper II). They found a good agreement between both
kinematical and morphological classifications. Only 16\% of galaxies
in this subsample are both kinematically-classified as RDs and
morphologically-classified as spirals. These ``rotating spiral disks''
were selected to have a redder bulge than the disk and to be therefore
quite similar to local spirals, apart their much bluer integrated
colors and [OII] equivalent widths (see Paper II for a detailed
discussion). Furthermore, they showed no special trend in size nor in
Bulge-to-Total luminosity ratio [$B/T$] compared to local galaxies,
and we do not expect any bias in the distant TFR that could be due to
morphological variation of the TFR along the Hubble Sequence (see,
e.g., \citealt{russell04}). In the following, we use the inclinations
[$i$] and half light radii [$R_{half}$] derived in Paper II.

The comparison between kinematics and morphology revealed two special
cases of RDs. First, J033230.78-275455.0 was detected in emission only
on one half of the galaxy, which was caused by the superposition of a
skyline on an [OII] emission line. This galaxy was classified as a RD,
but the amplitude of its VF is affected by large uncertainty. Second,
J033241.88-274853.9 was classified as a Tadpole galaxy in Paper II
because of its highly-asymmetric shape. In the following (see Sect.
3.2), we assume that all RD flux distributions can be approximated by
an exponential disk, which, obviously, does not apply to this galaxy.
Therefore, in this paper, ``RDs'' refers to galaxies classified as
rotating disks in Paper I \emph{excluding} J033230.78-275455.0 and
J033241.88-274853.9, while ``RD+'' refers to all RDs \emph{including}
J033230.78-275455.0 and J033241.88-274853.9.

\subsubsection{Absolute K-band magnitudes}
Absolute K-band magnitudes $M_K$ were taken from \cite{flores06} and
\cite{ravi07}. They were derived using Bruzual \& Charlot 2001
stellar-population models, assuming a CSP template with solar
metallicity and an exponential star-formation history with $\tau=0.5$
Gyr, which describes the properties of most galaxies between $z=0.4$
and 1 \citep{hammer01}. For each galaxy, an optimal SED was found by
matching the observed J-K color. This method was preferred to complete
SED fitting because 50\% of the selected galaxies, from the CFRS and
the HDFS, did not have similar multiwavelength photometric data to
galaxies in the remaining 50\%, which were selected from the CDFS; SED
fitting would, therefore, not have been able to measure an $M_K$ of
similar quality for all galaxies studied. We compared, for galaxies
from the CDFS, the impact of this simple ``color-matching'' method
relatively to a full SED fitting (using optical photometry only): we
found an excellent agreement between J-band absolute magnitudes with a
$1-\sigma$ dispersion $\sim0.01$ mag, and no systematic effect (Hammer
\& Pozzetti, private communication). We compared absolute J-band and
not K-band magnitudes because whatever the method used, these are not
extrapolations. A potential drawback of the ``color-matching'' method
is that $M_K$ is extrapolated, since the reddest photometric point
used is the observed K-band, which falls roughly in the middle of the
rest-frame J and H band at $z\sim0.6$. For galaxies in the CDFS, IRAC
$3.6 \mu$m fluxes were publicly available, which allowed us to compare
the $M_K$ obtained using this method with those obtained taking into
account the IR $3.6 \mu$m flux: we found a residual $1-\sigma$
dispersion of 0.12 mag between both estimates, with no systematic
effect. Using two independent photometric datasets for galaxies in the
CDFS (i.e., EIS and ACS), we derived a random uncertainty of $\lesssim
0.2$ mag on $M_K$, with no noticeable systematic uncertainties.
Therefore, we adopted a random uncertainty of 0.2 mag and neglected
systematic uncertainties.

Absolute magnitudes were corrected for internal extinction, using the
mass-dependent method of \cite{tully98}. We applied an additional
correction of 0.04 mag that accounted for extinction in face-on
galaxies \citep{tully98,tully85}. We note that within the sample of 68
objects, three do not have NIR photometry (see Table \ref{ech}).

\subsubsection{Summary}
For the distant sample, all measurements used are listed in Table
\ref{ech}. The columns of the table are as follows: (1) IAU ID, (2)
Internal ID, (3) Redshift $z$, (4) Inclination $i$ (in deg., $\pm5$
deg, see Sect. 2.1.2), (5) Dynamical Class D.C., (6) Projected radial
velocity $\Delta V_{obs}\sin{(i)}$ (in km/s, see Sect. 3.1), (7)
Rotation velocity corrected for inclination and spatial-resolution
effects $V_{flat}$ (in km/s, see Sect. 3.1), (8) Total uncertainty in
$V_{flat}$ (in km/s, see Sect. 3.4), (9) Absolute K-band magnitude
$M_K$ (uncorrected for extinction), (10) Absolute K-band magnitude
$M_K^c$ corrected for extinction ($\pm0.2$ mag, see Sect. 2.1.3), and
(11) Stellar mass in $\log{(M_{stellar}/M_\odot)}$ ($\pm0.3$ dex, see
Appendix A).

\begin{table*}[p]
\centering
\caption{Principle properties of the sample of 68 galaxies used in
  this study, ordered by increasing $RA$ (see text).}
\begin{tabular}{ccccccccccc}\hline
IAU ID & Internal ID & $z$ & $i$ & D.C. & $\Delta V_{obs}\sin{(i)}$ & $V_{flat}$ & $\Delta V_{flat}$ & $M_K$ & $M_K^{c}$ & $\log{(M_{stellar}/M_\odot)}$\\\hline

J030225.28+001325.1 & CFRS030085   & 0.6100 &  71 & RD  & 127 & 170 &  23 &  -20.83  & -20.98 & 10.21  \\
J030228.72+001333.9 & CFRS030046   & 0.5120 &  66 & RD  & 148 & 200 &  29 &  -21.68  & -21.82 & 10.64  \\
J030232.16+000639.1 & CFRS039003   & 0.6189 &  29 & RD  & 110 & 290 &  79 &   99.99  &  99.99 & 99.99  \\
J030238.74+000611.5 & CFRS031032   & 0.6180 &  37 & PR  &  83 & 200 &  36 &  -22.64  & -22.70 & 11.00  \\
J030239.38+001327.1 & CFRS030523   & 0.6508 &  41 & CK  &  63 & 130 &  38 &  -21.55  & -21.61 & 10.48  \\
J030240.45+001359.4 & CFRS030508   & 0.4642 &  38 & CK  &  47 & 100 &  40 &  -20.34  & -20.40 & 10.00  \\
J030240.99+000655.4 & CFRS031016   & 0.7054 &  67 & CK  &  15 &  30 &  25 &  -21.24  & -21.25 & 10.37  \\
J030242.19+001324.3 & CFRS030488   & 0.6069 &  41 & CK  &  36 &  70 &  36 &  -20.84  & -20.89 & 10.20  \\
J030245.67+001027.9 & CFRS030645   & 0.5275 &  45 & CK  &  89 & 150 &  38 &  -21.34  & -21.42 & 10.40  \\
J030246.94+001032.6 & CFRS030619   & 0.4854 &  27 & RD  &  70 & 210 &  76 &  -21.93  & -21.99 & 10.63  \\
J030248.41+000916.5 & CFRS031353   & 0.6340 &  57 & RD  & 174 & 260 &  41 &  -22.42  & -22.54 & 10.85  \\
J030249.10+001002.1 & CFRS031349   & 0.6155 &  48 & PR  & 175 & 290 &  46 &  -22.92  & -23.01 & 11.04  \\
J030252.03+001033.4 & CFRS031309   & 0.6170 &  71 & CK  &  84 & 150 &  22 &  -22.91  & -23.05 & 11.03  \\\hline
J033210.25-274819.5 & CDFS4301297      & 0.6100 &  69 & PR  & 110 & 150 &  26 &  -20.83  & -20.96 & 10.29  \\
J033210.76-274234.6 & CDFS4402679      & 0.4180 &  26 & CK  & 186 & 550 & 123 &  -23.62  & -23.68 & 11.44  \\
J033212.39-274353.6 & CDFS3400803      & 0.4230 &  90 & RD  & 105 & 180 &  22 &  -21.55  & -21.85 & 10.61  \\
J033213.06-274204.8 & CDFS3500001      & 0.4220 &  79 & CK  &  90 & 130 &  22 &  -20.56  & -20.73 & 10.19  \\
J033214.97-275005.5 & CDFS3300063      & 0.6680 &  22 & PR  &  62 & 190 &  90 &  -22.45  & -22.50 & 10.92  \\
J033217.62-274257.4 & CDFS3401109      & 0.6470 &  46 & CK  & 131 & 250 &  43 &  -21.15  & -21.24 & 10.38  \\
J033219.32-274514.0 & CDFS3400329      & 0.7250 &  72 & CK  & 116 & 270 &  29 &  -21.16  & -21.36 & 10.39  \\
J033219.61-274831.0 & CDFS3300651      & 0.6710 &  49 & PR  &  82 & 190 &  33 &  -20.90  & -20.98 & 10.27  \\
J033219.68-275023.6 & CDFS3202670      & 0.5610 &  58 & RD  & 161 & 230 &  33 &  -22.30  & -21.42 & 10.88  \\
J033220.48-275143.9 & CDFS3202141      & 0.6790 &  63 & CK  &  39 &  70 &  24 &  -20.63  & -20.70 & 10.18  \\
J033224.60-274428.1 & CDFS3400618      & 0.5380 &  65 & CK  &  63 &  90 &  27 &  -20.35  & -20.44 & 10.08  \\
J033225.26-274524.0 & CDFS3400279      & 0.6660 &  60 & CK  &  47 &  80 &  26 &  -21.54  & -21.61 & 10.56  \\
J033226.23-274222.8 & CDFS3401338      & 0.6679 &  76 & PR  & 150 & 200 &  24 &  -21.93  & -22.13 & 10.72  \\
J033227.07-274404.7 & CDFS3400743      & 0.7390 &  84 & CK  &  71 & 110 &  21 &  -20.95  & -21.15 & 10.26  \\
J033228.48-274826.6 & CDFS3300684      & 0.6697 &  22 & CK  &  23 & 130 &  66 &  -21.68  & -21.73 & 10.63  \\
J033230.43-275304.0 & CDFS2200433      & 0.6460 &  70 & CK  & 219 & 380 &  29 &  -21.68  & -21.90 & 10.64  \\
J033230.57-274518.2 & CDFS2400243      & 0.6810 &  34 & CK  &  41 &  80 &  46 &  -22.86  & -22.91 & 11.08  \\
J033230.78-275455.0 & CDFS2102060      & 0.6870 &  66 & RD+ &  90 & 200 &  43 &  -21.82  & -21.96 & 10.66  \\
J033231.58-274121.6 & CDFS2500322      & 0.7047 &  42 & RD  &  81 & 140 &  41 &  -20.51  & -20.57 & 10.10  \\
J033232.96-274106.8 & CDFS2500425      & 0.4690 &  16 & PR  &  31 & 210 & 117 &  -20.13  & -20.17 & 10.01  \\
J033233.90-274237.9 & CDFS2401349      & 0.6190 &  17 & PR  &  46 & 200 & 107 &  -21.82  & -21.96 & 10.66  \\
J033234.04-275009.7 & CDFS2300055      & 0.7030 &  59 & RD  & 104 & 160 &  29 &  -20.50  & -20.60 & 10.09  \\
J033234.12-273953.5 & CDFS2500971      & 0.6280 &  32 & CK  &  25 & 110 &  44 &   99.99  &  99.99 & 99.99  \\
J033237.54-274838.9 & CDFS2300477      & 0.6650 &  31 & RD  & 106 & 230 &  71 &  -21.98  & -22.03 & 10.70  \\
J033238.60-274631.4 & CDFS2301047      & 0.6220 &  60 & RD  & 118 & 210 &  33 &  -21.45  & -21.57 & 10.53  \\
J033239.04-274132.4 & CDFS2500233      & 0.7330 &  43 & PR  &  58 & 130 &  39 &  -20.62  & -20.68 & 10.14  \\
J033239.72-275154.7 & CDFS2200829      & 0.4160 &  35 & CK  &  16 &  30 &  36 &  -20.94  & -20.98 & 10.31  \\
J033240.04-274418.6 & CDFS2400536      & 0.5223 &  16 & CK  & 108 & 470 & 214 &  -21.95  & -22.00 & 10.77  \\
J033241.88-274853.9 & CDFS2300404      & 0.6680 &  67 & RD+ &  80 & 120 &  27 &  -20.90  & -21.01 & 10.27  \\
J033243.62-275232.6 & CDFS2200611      & 0.6800 &  71 & PR  &  38 &  60 &  23 &  -19.94  & -20.01 &  9.86  \\
J033244.20-274733.5 & CDFS2300750      & 0.7365 &  39 & CK  &  46 & 170 &  40 &  -21.76  & -21.83 & 10.62  \\
J033245.11-274724.0 & CDFS2300800      & 0.4360 &  43 & RD  &  95 & 270 &  44 &  -22.03  & -22.11 & 10.80  \\
J033248.28-275028.9 & CDFS1202537      & 0.4462 &  81 & PR  &  76 & 110 &  22 &  -20.38  & -20.54 & 10.09  \\
J033249.53-274630.0 & CDFS1302369      & 0.5230 &  46 & PR  &  74 & 150 &  39 &  -21.00  & -21.07 & 10.34  \\
J033250.24-274538.9 & CDFS1400714      & 0.7318 &  31 & CK  &  82 & 240 &  67 &  -20.59  & -20.65 & 10.12  \\
J033250.53-274800.7 & CDFS1301018      & 0.7370 &  62 & PR  &  71 & 110 &  26 &  -20.34  & -20.43 & 10.01  \\\hline
J221741.46+001854.8 & CFRS221119   & 0.5138 &  41 & RD  & 135 & 250 &  47 &   99.99  &  99.99 & 99.99  \\
J221743.08+001508.3 & CFRS221064   & 0.5383 &  48 & PR  &  93 & 250 &  29 &  -21.64  & -22.74 & 10.53  \\
J221745.12+001447.4 & CFRS220975   & 0.4211 &  50 & CK  & 313 & 410 &  44 &  -22.53  & -22.64 & 10.95  \\
J221746.48+001653.5 & CFRS220919   & 0.4738 &  60 & CK  &  39 &  30 &  23 &  -19.53  & -19.55 &  9.67  \\
J221754.56+001900.3 & CFRS220619   & 0.4676 &  68 & PR  &  62 &  90 &  24 &  -19.33  & -19.42 &  9.63  \\
J221758.07+002137.5 & CFRS220504   & 0.5379 &  42 & RD  &  93 & 170 &  42 &  -21.36  & -21.43 & 10.43  \\
J221802.92+001428.0 & CFRS220321   & 0.4230 &  42 & PR  & 104 & 220 &  46 &  -20.79  & -20.87 & 10.21  \\
J221803.55+002131.9 & CFRS220293   & 0.5420 &  45 & CK  &  89 & 160 &  40 &  -20.89  & -20.97 & 10.24  \\\hline
J223241.45-603516.1 & HDFS4130     & 0.4054 &  36 & CK  &  90 & 220 &  53 &  -22.13  & -22.20 & 10.76  \\
J223245.56-603418.8 & HDFS4170     & 0.4602 &  51 & RD  & 134 & 230 &  37 &  -22.60  & -22.70 & 10.96  \\
J223252.74-603207.3 & HDFS4040     & 0.4650 &  51 & PR  &  75 & 120 &  32 &  -20.04  & -20.11 &  9.88  \\
J223254.05-603251.6 & HDFS4090     & 0.5162 &  45 & CK  &  14 &  20 &  30 &  -19.83  & -19.85 &  9.80  \\
J223256.07-603148.8 & HDFS4020     & 0.5138 &  50 & RD  &  88 & 140 &  36 &  -20.12  & -20.20 &  9.96  \\
J223256.08-603414.1 & HDFS5140     & 0.5649 &  50 & CK  & 171 & 220 &  42 &  -20.46  & -20.56 & 10.04  \\
J223257.52-603305.9 & HDFS5030     & 0.5821 &  25 & CK  &  31 & 130 &  69 &  -22.68  & -22.73 & 10.94  \\
J223258.01-603525.9 & HDFS4180     & 0.4647 &  64 & RD  & 104 & 140 &  26 &  -20.38  & -20.49 & 10.04  \\
J223258.23-603331.4 & HDFS4070     & 0.4230 &  40 & CK  &  36 &  70 &  43 &  -19.67  & -19.72 &  9.75  \\
J223300.09-603529.9 & HDFS5190     & 0.6952 &  59 & RD  & 144 & 230 &  23 &  -21.92  & -22.04 & 10.64  \\
J223302.45-603346.5 & HDFS5150     & 0.6956 &  42 & PR  &  51 & 120 &  39 &  -21.02  & -21.09 & 10.27  \\\hline
\end{tabular}
\label{ech}
\end{table*}

\subsection{Local sample}
As a local reference, we adopt the K-band TFR derived by
\cite{hammer07} for a complete subsample of the SDSS
\citep{pizagno07}, which allows us to control systematic effects that
can occur when comparing local and distant samples.

One important choice for studying the TFR is the kinematical estimator
used for the rotation velocity $V_{rot}$. Studies of the local TFR
have shown that using different estimators (e.g., the maximal rotation
velocity $V_{max}$, the plateau rotation curve velocity $V_{flat}$,
the velocity measured at the radius containing 80\% of the light
$V_{80}$, and the velocity $V_{2.2}$ measured at 2.2 disk scale
length) can lead to different results (see, e.g.,
\citealt{ver01,pizagno07}). $V_{flat}$ has been shown to be correlated
with $V_{rot}$, since it is less influenced by the bulge dynamics,
which can produce a central ``bump'' with $V_{max}>V_{flat}$. This can
then lead to a tighter TF relation (e.g., \citealt{ver01}), and a
significant improvement in the linearity of the K-band relation at the
high-mass end \citep{noordermeer07}. For the local sample, we adopted
the $V_{80}$ measurements of \citep{pizagno07} using $\arctan$ fits to
the RCs. To limit uncertainties, we restricted the local sample to
galaxies for which $V_{80}$ is a good proxy for $V_{flat}$ (i.e.,
\citealt{pizagno07} flags 1 and 2), as shown by \cite{hammer07}.

Rotation velocities were corrected for inclination using estimates
derived from their morphological axis ratio. When well-resolved 2D
kinematics is available, it is possible to derive the inclination
directly from the fit of the VF. However, large differences (up to
$\sim10$ deg) can be found between such kinematically-derived
inclinations and the ones inferred from the morphological axis ratio
(e.g., \citealt{chemin06}). Since we do not have 2D kinematics for
galaxies in the local sample, we use morphologically-derived
inclinations exclusively. This provides us with homogeneous estimates
for the local and distant samples, since in the latter we also use
such inclinations.

\cite{hammer07} combined the \cite{pizagno07} kinematic measurements
with 2MASS Ks-photometry. We estimated absolute magnitudes in the
local sample following a similar method to that used for the distant
sample, including corrections for extinctions (see Sect. 2.1.3). The
2MASS Ks filter has the advantage of being close to the ISAAC Ks
filter used in the distant sample. Both filters match well the K-band
LCO filter designed to establish the faint IR standard-star system of
\cite{persson98} \citep{mason07,carpenter01}. We assumed that
both filters are identical during the SED-fitting procedure, which
did not introduce any systematic effect. We note than 2MASS K-band
magnitudes were converted into AB magnitudes using
$Ks(AB)=Ks(Vega)+1.85$, following \cite{blanton05}.

Since we are exploring the higher tail of the stellar-mass
distribution and the TFR is highly sensitive to stellar mass (see,
e.g., \citealt{McGaugh05}), we restrict the K-band TFR to
$\log{(V_{flat})} \geq 2.2$ and find from \cite{hammer07}:
\begin{equation}
$$M_K(AB)=-6.54\pm1.33-(6.88\pm0.57)\times \log{(V_{flat})},
\label{eqmk}
\end{equation}
 with a residual dispersion $\sigma _{res}=0.38$ mag. Since we used
 uniform $M_K$ error bars, we used direct fits to the TFR, i.e., with
 $M_K$ as a $V_{flat}$-dependent variable (using the IDL MPFITFUN
 procedure of C. Markwardt, translated from the MINPACK-1
 package\footnote{http://cow.physics.wisc.edu/$\sim$craigm/idl/mpfittut.html}).

\section{Methodology}

\subsection{Derivation of the rotation velocity}
For each galaxy, we estimated the deprojected VF half-amplitude using
$\Delta V _{obs}=(V_{max}^{IFU}-V_{min}^{IFU})/(2\sin{(i)})$, where
$V_{max}^{IFU}$ and $V_{min}^{IFU}$ are respectively the maximal and
minimal values of the VF sampled by the IFU (see Table \ref{ech}).

Because of the influence of the relatively coarse spatial-resolution
of the kinematic data, $\Delta V _{obs}$ underestimates the true
rotation velocity $V_{rot}$ (see \citealt{flores06,puech06}). Rather
than applying a mean correction factor to the entire sample as in
\cite{flores06}, we corrected ech galaxy \emph{individually} by
modeling its data-cube. We used a method developed from that used by
\cite{flores06} and in Paper I, to model the $\sigma$-map. Assuming
that all galaxies are thin-rotating disks, we modeled their data-cube,
from their observed VF and $\sigma$-map, in the following way.

\begin{figure}[h!]
\centering
\includegraphics[width=9cm]{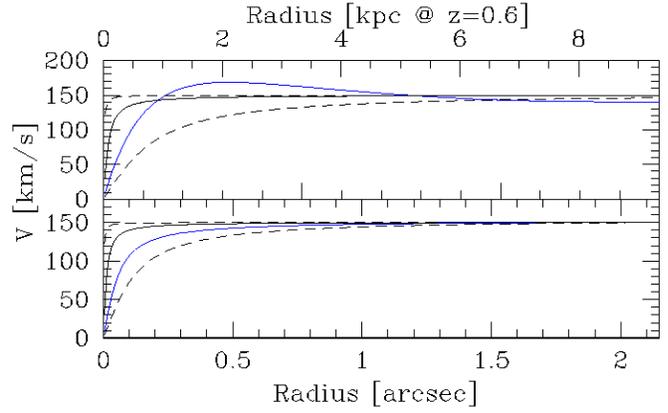}
\caption{Illustration of the $\Delta V _{obs}$ correction method, for
  an early-type RC (\emph{upper panel}, note that $V_{max}>V_{flat}$)
  and a late-type RC (\emph{bottom panel}). The blue thick curves
  represent the input RC of the test, the black curves the best models
  (see text), and the dash-lines the two alternative RCs with
  non-optimal $r_t$. We note that typical $r_t$ values lead to
  velocity gradients on spatial scales that are much smaller than the
  typical seeing of $\sim0.8$ arcsec.}
\label{test}
\end{figure}

First, for each galaxy, we constructed a grid of rotation curves (RC)
with $V_{rot}$ spaced at 10 km/s intervals, which roughly corresponded
to the typical uncertainty in $\Delta V _{obs}$ (see Sect. 3.4).
Because of a lack of spatial resolution, the precise shape of distant
RCs remains largely unknown. Therefore, we chose a simple $\arctan$
defined as $V_0\times (2/\pi)\times \arctan{(r/r_t)}$, which depends
only on two parameters, i.e., the asymptotic velocity $V_0$, and the
``turnover'' radius $r_t$ (see, e.g., \citealt{courteau97}). This RC
shape was used in Paper I to model the VF of each distant galaxy. In
the RD+ subsample, such models provided good matches to the observed
VFs, which demonstrated that such a RC shape is a reasonable choice.
The same shape was adopted in the local sample (see Sect. 2.2), which
provided us with homogeneous estimates of the rotation velocity in
both the local and distant samples. We chose not to fit a wide range
of values for $r_t$ because a visual inspection of VFs revealed that
for almost all RDs, the gradient of the RC fell inside a single
GIRAFFE IFU pixel (\citealt{flores06}, Paper I). We therefore explored
only three fixed values for $r_t$ (see Fig. \ref{test}), which allowed
us to investigate more extreme cases where the RC is relatively steep
or, in contrast, relatively flat.

The $\arctan$ RC model cannot reproduce the central ``bump'' observed
in some early-type galaxies (see, e.g., Fig. \ref{test}). Given the
coarse spatial resolution of the GIRAFFE IFU (0.52 arcsec/pix, i.e.,
$\sim3.5$ kpc at $z=0.6$) and the large size of the seeing disk
compared to the typical scalelengths of distant galaxies, it might
however be impossible to distinguish clearly between $V_{max}$ or
$V_{flat}$ in distant galaxies. It might be expected that the
asymptotic velocity $V_0$, corresponding to an $\arctan$ RC, is
probably, in this case, an average value between $V_{max}$ and
$V_{flat}$. Because we rely on $V_{flat}$ as a kinematic measure of
the rotation velocity in the local sample (see Sect. 2.2), we are
naturally led to quantify all random and systematic uncertainties
relative to $V_{flat}$. Finally, we note that even for late-type RCs,
$V_0 \sim V_{flat}$ only if the RC has a well-defined plateau and if
this plateau can be observed in terms of spatial coverage or SNR.

From each RC, a high-resolution data-cube was constructed, assuming a
simple Gaussian shape for the emission line. For simplicity, we did
not include noise in the simulations, assuming that the global
uncertainty can be derived by estimating separately the influence of
the noise and other effects (see Sect. 3.4). As a consequence,
neglecting the doublet spectral nature of the [OII] emission line
should not impact severally the results. The velocity width was
assumed to be the minimal value observed in the $\sigma$-map, and the
emission-line flux was taken from an exponential flux distribution,
assuming $R_d\sim R_{half}/1.6$ \citep{persic91} and limiting the
spatial extent to a radius $2R_{half}\sim 3R_d\sim R_{opt}$, where
$R_d$ is the disk scale-length, and $R_{opt}$ is the optical radius.
We chose to use $R_{half}$ rather than direct $R_{d}$ measurements
because more than 80\% of galaxies in the sample were not classified
as simple spirals but had more complex morphologies (see Paper II),
which could lead to meaningless $R_d$ values. In contrast, $R_{half}$
provides us with a uniform and well-defined size-parameter, which can
be safely converted into $R_d$ for thin exponential disks: using the
bulge/disk decomposition of the RD subsample done in Paper II, we
compared $R_d$ with $R_{half}/1.6$, and found a \emph{maximal}
difference of $\sim0.1$ arcsec, which is much smaller than the typical
seeing during the observations (see also \citealt{puech07}). Thus,
such an error in $R_d$ has little influence on the model, and we
chose not to explore this parameter.

\begin{figure}[h!]
\centering
\includegraphics[width=9cm]{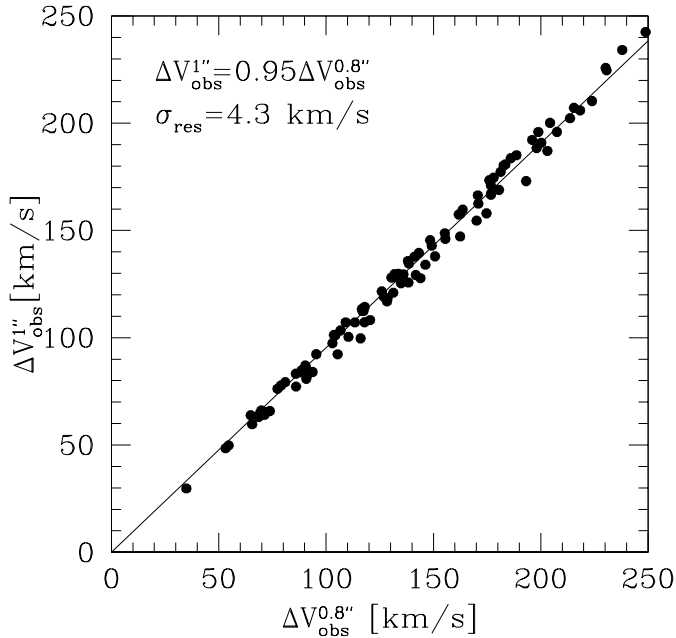}
\caption{Comparison between the $\Delta V_{model}$ values obtained in
  Monte-Carlo simulations of 100 GIRAFFE data-cubes using a 0.8 and 1
  arcsec seeing (see text). The black line is a linear fit fixing the
  intercept to zero. The residual dispersion is $\sim4.3$ km/s.}
\label{Figseeing}
\end{figure}

Each high-resolution data-cube was convolved by a 0.8 arcsec seeing,
which corresponded to the median condition of the observations, and
then rebinned to the GIRAFFE sampling, i.e., 0.52 arcsec/pix. From
these simulated GIRAFFE data-cubes, simulated VFs and simulated
$\Delta V_{model}$ were derived as for real GIRAFFE data. We checked
the influence of changing the seeing from 0.8 to 1.0 arcsec, on the
$\Delta V_{model}$: using Monte-Carlo simulations of 100 GIRAFFE
data-cubes (see next section), we found a good linear correlation
between the $\Delta V_{model}$ obtained for a 1 arcsec seeing and that
obtained for a 0.8 arcsec seeing, all other properties being equal.
This fit indicated that for a 1 arcsec seeing, $\Delta V_{model}$ was
reduced by $\sim0.05$\% compared to that obtained for a 0.8 arcsec
seeing (see Fig. \ref{Figseeing}). Therefore, the maximal uncertainty
on $\Delta V_{model}$ due to seeing variations is $\sim12$ km/s, which
corresponds roughly to the velocity spacing adopted for the searching
grid used to correct $\Delta V_{obs}$ (see above), which implies that
this velocity grid was well adapted for our purpose.

\begin{figure*}
\centering
\includegraphics[width=18cm]{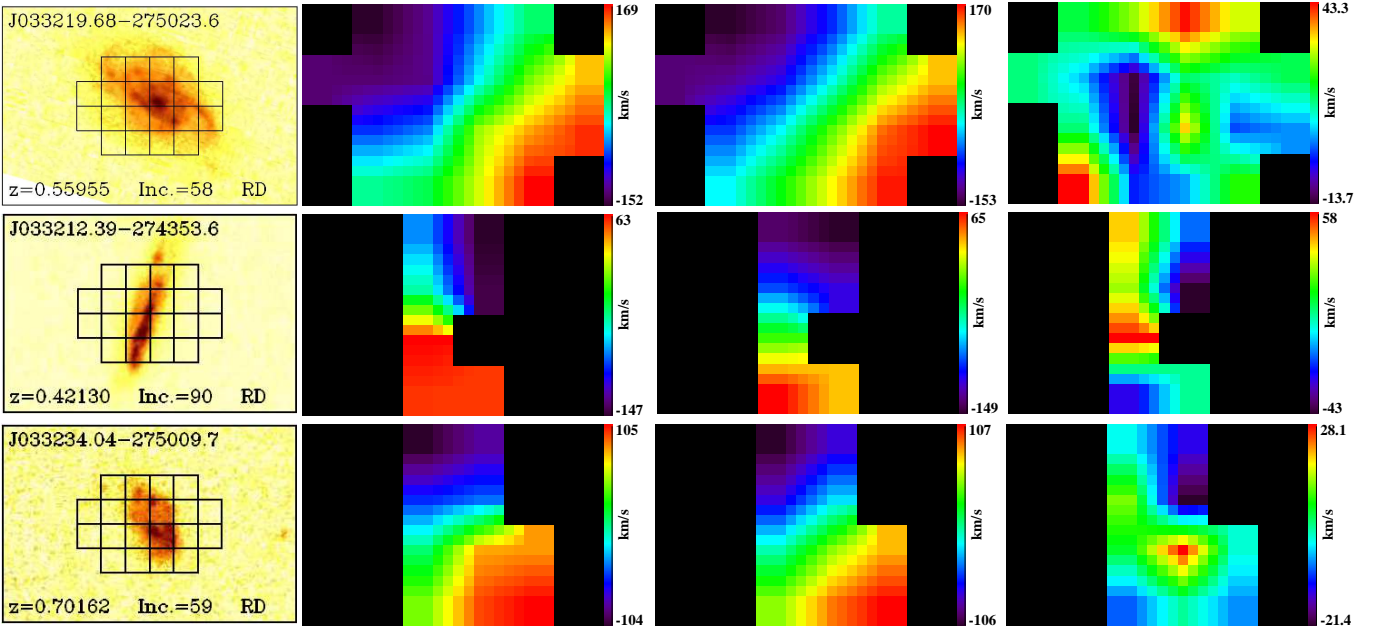}
\caption{Examples of kinematical fitting of three $z\sim0.6$ RD
galaxies. \emph{From left to right:} HST/ACS F775W image with the
GIRAFFE IFU superimposed (from Paper I), observed VF (shown with a 5x5
interpolation see Paper I), best modeled VF (shown with a 5x5
interpolation), residual map between the observed and modeled VFs.
Relatively large differences are found only close to the minor axis,
where departure from pure circular motion is artificially exaggerated
by projection effects (see a discussion of this effect in, e.g.,
\citealt{chemin06}).}
\label{fitex}
\end{figure*}

To find the best model, we finally looked for the ($r_t$,$V_{flat}$)
pair minimizing the difference between $\Delta V_{model}$ and $\Delta
V_{obs}$. We checked that, in all simulations, such a criterion gives
a unique solution. Results are listed in Table \ref{ech}, and
representative examples of this kinematical fitting for three RD
galaxies are shown in Fig. \ref{fitex}.

\subsection{Testing the method of correction}
To test the correction method, we performed Monte-Carlo simulations of
100 GIRAFFE data-cubes. The parameter space investigated encompassed
the inclination, half-light radius, $PA$, rotation velocity, and RC
gradient, of values randomly chosen from the typical values observed
for the GIRAFFE sample. Half of the Monte-Carlo simulations were
generated using an $\arctan$ RC shape model, and the other half using
an exponential term combined with a second-order polynomial term. This
polynomial term allowed us to create a central velocity ``bump'' as
observed in some early-type galaxies (see Fig. \ref{test}). We note,
for late-type RCs, that the $V_{flat}$ did not have to be located
within the IFU, which allowed us to test how the correction procedure
operated in this case. For early-type galaxies, $V_{flat}$ should be
located within the IFU, as $V_{max}$ is generally located close to
$2.2R_d\sim1.375R_{half}$ (e.g., \citealt{courteau97}), which almost
always falls within the IFU \citep{flores06}. We note, however, that
in some cases, because of the dynamical influence of the bulge, the RC
of some early-type galaxies can show an extended velocity peak up to
large distances (see, e.g., M31, \citealt{carignan06}). The limited
spatial coverage of the GIRAFFE IFU would probably provide us with an
overestimation of the rotation velocity. In the RD sample, all
galaxies besides one have $B/T<0.2$, which means that such an effect
cannot affect significantly the RD sample.

\begin{figure}[h!]
\centering
\includegraphics[width=9cm]{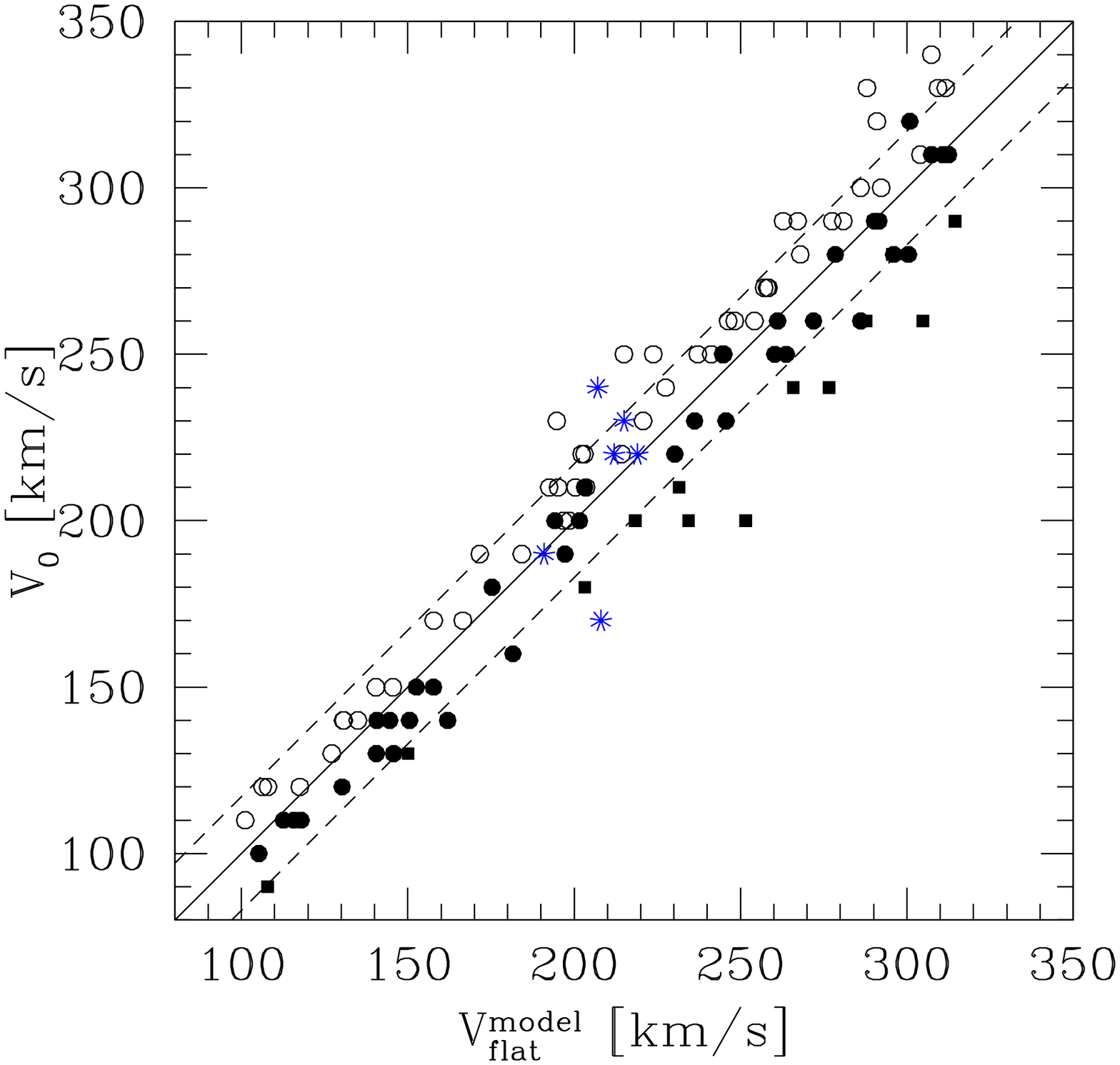}
\caption{Comparison between the $V_{flat}$ used as inputs to the
  Monte-Carlo simulations of 100 GIRAFFE data-cubes, with the $V_0$
  values obtained after using the method of correction described in
  Sect. 3.2. Black dots represent RCs generated using an $\arctan$
  model (i.e., late-type like RCs), while open circles represent RCs
  typical of early-type galaxies. Black squares represent simulations
  where the rotation velocity is not sampled by the IFU (see text).
  The black line is a linear fit, and the dash lines represent the
  $1-\sigma$ residual dispersion $\sim$17 km/s. Blue stars represent
  real observations of local galaxies artificially redshifted to
  $z\sim0.6$ (see text).}
\label{compv}
\end{figure}

These simulated RCs were used to simulate GIRAFFE observations
following the method outlined in Sect. 3.2. These simulated data-cubes
were in turn used as inputs to test the method of correction on
$\Delta V_{obs}$. In Fig. \ref{compv}, we compare the input rotation
velocity $V_{flat}$, with the asymptotic velocity $V_0$ of the best
$\arctan$ model obtained using the method detailed above. We find a
good linear correlation with
$V_0=[0.65\pm6.22]+[1.00\pm0.03]V_{flat}$, consistent with
$V_0=V_{flat}$, and a residual dispersion $\sim17$ km/s. In this
figure, we have distinguished between simulations where $V_{flat}$ is
sampled by the IFU (black circles) from those where this is not the
case (black squares). If we compare the corrected velocity $V_0$ with
the last point of the RC sampled by the IFU $V_{end}$, both sets of
simulations fall on the same region. In such a plot (not shown here),
we find a similar result, with
$V_0=[-1.23\pm4.47]+[1.03\pm0.02]V_{end}$, also consistent with
$V_0=V_{end}$.

We checked that $r_t$ had little influence on the $\Delta V_{obs}$
correction by comparing the corrected rotation velocity, derived by
allowing $r_t$ to vary between the three values defined in Sect. 3.1,
with that obtained by fixing $r_t$ to the middle value of this grid,
i.e. the one that makes most of the RC gradient fell inside one
GIRAFFE pixel as observed in most cases. We found that both sets of
values correlated well within $\sim$ 6 km/s ($1-\sigma$), i.e., as
expected, $r_t$ has a negligible impact on the derived correction.
This is due mainly to the large seeing size ($\sim$0.8 arcsec)
compared to velocity gradients associated with typical $r_t$ values
(see Fig. \ref{test}), which significantly dilutes the variations of
the RC gradient, once projected onto the IFU.

It is important to check whether or not these simulations are
representative of real galaxies. To do this, we selected a few
galaxies from the GHASP survey \citep{amram02}, which provides us with
high spatial-resolution data-cubes for a morphologically-complete
sample of local galaxies using Fabry-Perot interferometry. The
analysis of the GHASP sample is still underway \citep{Epinat08}, and a
full comparison between local 3D data and distant 3D GIRAFFE data will
be addressed in a future work. Here, we check whether or not the
simulations are representative of local galaxies, especially in the
range of velocity and size spanned by distant RDs. We restrict our
choice to galaxies for which an $\arctan$ shape provides a relatively
good representation of the RC, because we probe the accuracy of the
correction applied to $\Delta V_{obs}$, and not to the measurement
itself, which is an independent issue. Given the limited number of
galaxies meeting these criteria, we chose to restrict our choice to 7
such galaxies with rotation velocities ranging from 190 to 220 km/s,
where are typical measurements for most distant RDs (see next Sect.
4). This choice allows us to sample roughly the most relevant velocity
range for distant RDs, rather than testing a few isolated points
spread over the full velocity range of the distant sample. We
simulated GIRAFFE observations by degrading the Fabry-Perot
high-resolution data-cube to the resolution of GIRAFFE observation
(0.8 arcsec seeing) and then to the GIRAFFE IFU spatial sampling. From
these data-cubes, we extracted a VF and $\Delta V_{obs}$ as for real
GIRAFFE data, and corrected them. These simulation results are shown
as blue stars datapoints in Fig. \ref{compv}, which agree with the
Monte-Carlo simulations that fall in the same velocity range,
reproducing the dispersion of the correlation. This confirms that we
can confidently rely on the correction applied to $\Delta V_{obs}$.

\begin{figure}[h]
\centering \includegraphics[width=8cm]{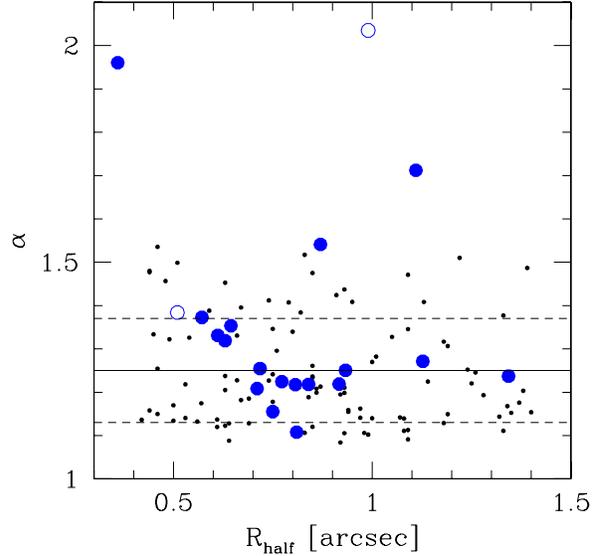}
\caption{Correction factors $\alpha$ used to correct $\Delta V _{obs}$
  for RDs (blue dots; open blue dots represent the two RD+ galaxies).
  Small black dots correspond to Monte-Carlo simulations of 100
  GIRAFFE data-cubes in the same range of half-light radius, PA,
  inclination, rotation velocity, and RC gradient (see text). The
  horizontal dash line represents to the mean correcting factor of
  1.25 derived from Monte-Carlo simulations, while the $1-\sigma$
  dispersion around the mean is shown in dash lines.}
\label{alpha}
\end{figure}

In Fig. \ref{alpha}, we plot all correction factors $\alpha
=V_{flat}/\Delta V _{obs}$ obtained for the Monte-Carlo simulations.
Both simulations of redshifted local galaxies and real GIRAFFE
rotating disk galaxies fall in the same region of the plot, appart
from three galaxies (J033212.39-274353.6,
J033230.78-275455.0,J033245.11-274724.0), that show relatively large
$\alpha$. One is a compact galaxy (J033245.11-274724.0, see Paper II),
while another one is seen almost edge-on (J033212.39-274353.6), which
might explain the relatively high $\alpha$. The last galaxy
(J033230.78-275455.0) is a special case, because only half of the
galaxy is detected in emission (this galaxy was shifted into the RD+
class in Sect. 2.1). We checked that the results presented in this
paper are not significantly affected when these three special objects
are removed from the sample. In the RD subsample, we found a median
$\alpha \sim$1.26, consistent with what is found in the Monte-Carlo
simulations, with a mean $\alpha=1.25\pm0.12$.

\subsection{Random uncertainty budget of the rotation velocity derivation}
The uncertainty budget can be decomposed into uncertainties related to
$\Delta V_{obs}$, and those related to the correction applied to
$\Delta V_{obs}$ to obtain $V_{flat}$.

\begin{figure}[h]
\centering
\includegraphics[width=9cm]{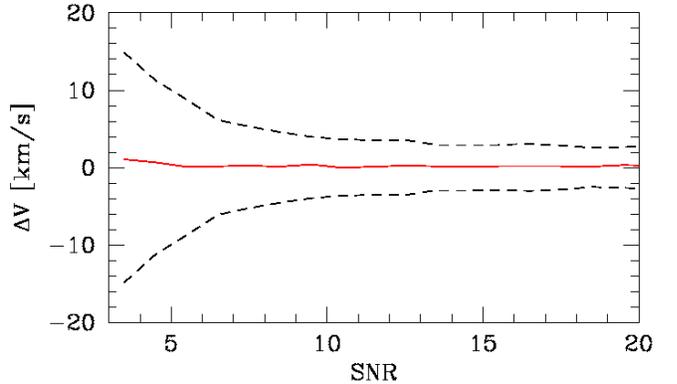}
\caption{Monte-Carlo simulations of the velocity measurement accuracy
  (see Paper I for details). The red line shows the corresponding bias
  (almost zero), while the black dashed lines shows the $1-\sigma$
  error: $\sim$12 km/s between SNR=3-5, $\sim$5 km/s between SNR=5-10,
  and $\leq$ 2 for SNR$\geq$10.}
\label{error}
\end{figure}

The main source of uncertainty that can affect $\Delta V_{obs}$ is
that associated with a finite spectroscopic SNR in the measurement of
$V_{max}^{IFU}$ and $V_{min}^{IFU}$ , which can be quantified using
Monte-Carlo simulations (see Fig. \ref{error}). For each galaxy, we
used the SNR maps derived in Paper I to estimate the mean SNR
uncertainty on $V_{max}^{IFU}$ and $V_{min}^{IFU}$, i.e., on $\Delta V
_{obs}$. We found a median (mean) uncertainty due to a finite SNR of
$\sim$9 km/s (8km/s).

We used Fig. \ref{compv} to quantify the uncertainty associated with
the correction made to $\Delta V_{obs}$. We found 17 km/s ($1-\sigma$
residual dispersion), with no noticeable systematic uncertainties. We
note that this uncertainty takes into account cases where the RC is
not fully sampled by the IFU. As we have already pointed out, one
galaxy in the sample deserves more consideration. Only part of the
galaxy J033230.78-275455.0, which was classified as RD+ (see Sect.
2.1), was detected in emission (see Paper I). $\Delta V _{obs}$ does
not encompass a symmetric range along the RC, which biases the
rotation-velocity input of the model to a lower value. Considering a
rough RC model, composed of a first, linear, rising part up to
$R_{max}=2.2R_D$ \citep{persic91}, and a second flat part, we estimate
a $\sim$30km/s correction to ``symmetrize'' $\Delta V _{obs}$. This
correction is convolved directly by the other uncertainty factors on
$\Delta V_{obs}$. Following Sect. 3.1 and 3.2, we also took into
account uncertainties due to seeing and $r_t$ variations ($\sim$5 and
6 km/s, respectively), which produced a total uncertainty for the
correction on $\Delta V_{obs}$ of $\sim$19 km/s.

The total uncertainty on $V_{flat}\times \sin{(i)}$ (i.e., the radial
rotation velocity uncorrected for inclination) was derived by adding
in quadrature all the previous terms. We finally propagate the
uncertainty associated with the inclination, which is $\pm$5 degrees,
according to Paper I. We finally found a median (mean) total
uncertainty on $V_{flat}$ of $\sim$37 km/s (44km/s). The $1-\sigma$
dispersion of the total uncertainties around the mean is $\sim$30 km/s.

\section{The K-band Tully-Fisher relation at $z\sim0.6$}

\subsection{Results}
\begin{figure}
\centering
\includegraphics[width=8.9cm]{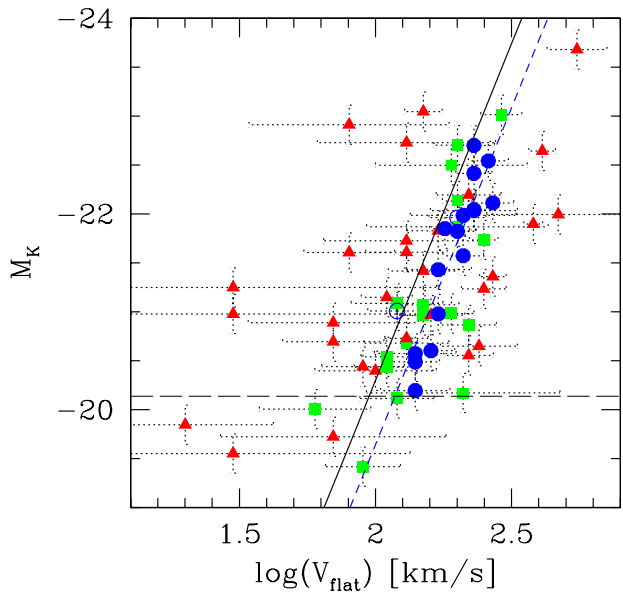}
\caption{Evolution of the K-band TFR ($AB$ magnitudes). The completude
  limit $M_K \sim-20.14$ (corresponding to $M_J=-20.3$, see Sect. 2)
  is indicated by an horizontal dash line. Blue dots represent RDs
  (the two RD+ galaxies are represented with open blue dots), green
  squares PRs, and red triangles CK galaxies. The black line is the
  local TFR, while the blue dash-line represent a linear fit to the
  $z\sim0.6$ TFR (see text).}
\label{tf}
\end{figure}

We compare both the distant and local K-band TFR in Fig. \ref{tf}.
Holding the slope to the local value, $\sigma _{res}$ increases from
RDs (0.31 mag), to PRs (0.80 mag), and CKs (2.08 mag). Restricting the
distant TFR to dynamically well-relaxed RDs, the local and distant
relations have comparable $\sigma _{res}$: we therefore confirm
quantitatively the previous findings of \cite{flores06} that all of
the enlarged dispersion of the distant TFR comes from non-relaxed
galaxies. If we allow the slope to vary during the fit, we find a
residual dispersion of 0.31 mag. This indicates that assuming no
evolution in slope appears to be a reasonable assumption.

The zero point of the TFR of distant RD galaxies is found to be
$-5.88\pm0.09$ ($1-\sigma$ bootstrapped error, slope fixed), i.e.,
0.66 mag fainter that the local zero point. Allowing the slope to vary
during the fit, we find a zero point of $-5.07\pm2.37$ and a slope of
$-7.24\pm1.04$ in agreement (within the corresponding uncertainties)
with those derived by fixing the slope to the local value. This
strongly suggests no evolution in slope of the K-band TFR. Given the
limited number of RDs in the distant sample (16), we adopt the
zero-point value derived by keeping the slope to the local value. We
note that if we consider the RD+ class (18 galaxies, see Sect. 2), we
find similar results (see Table \ref{fittfk}).

\begin{table*}
\centering
\caption{Fits to the local and distant K-band TFRs, using
  $M_K(AB)=a+b\times \log{(V_{flat})}$.}
\begin{tabular}{lccl}\hline\hline
K-band TFR        & Slope a        & Zero point b   & Comments         \\\hline
Local relation    & -6.88$\pm$0.57 & -6.54$\pm$1.33 & SDSS subsample  \\\hline
Distant RDs       & -6.88          & -5.88$\pm$0.09 & Using local slope\\
Distant RDs       & -7.24$\pm$1.04 & -5.07$\pm$2.37 & Slope free       \\\hline
Distant RD/RD+    & -6.88          & -5.92$\pm$0.10 & Using local slope\\
Distant RD/RD+    & -6.47$\pm$1.30 & -6.85$\pm$2.98 & Slope free       \\\hline\hline
\end{tabular}
\label{fittfk}
\end{table*}

We find in the distribution of $M_K$ residuals, a skewness and a
kurtosis of -0.22 and -0.99, respectively for RDs: this is roughly
consistent with the Gaussian residuals, within the $1-\sigma$ expected
thresholds\footnote{For a distribution of n points, these thresholds
are $\sqrt(6/n)$ and $\sqrt(21/n)$ respectively (n=16 is the present
case).}. This indicates that residuals are not biased significantly on
any side of the relation. A Welcher's t-test then provides a probability
$\ll$1\% that the local and distant relations have the same zero
point. We note that during the fitting procedure, we weighted all
the rotation velocities by their associated uncertainties. Therefore,
the result of the Welcher's t-test means that within random
uncertainties, the zero-point difference of 0.66 mag between the
distant and the local relations is statistically significant.

\subsection{Systematic uncertainty budget}
We investigate which systematic effects could bias the evolution
of TFR zero point detected between $z\sim0.6$ and $z=0$.

\subsubsection{Photometric systematic uncertainty}
As stated in Sect. 2, we do not expect any noticeable
systematic-effect on measurements due to photometric measurement or
calibration. However, one possible exception is the extinction
correction made on both local and distant galaxies. Even if we follow
exactly the same procedure for the two samples, one might wonder
whether or not the low surface-brightness outer-regions in distant
galaxies could be far less well detected compared to local galaxies,
which would create a bias in the derivation of their axis ratios and
extinction corrections. However, we found in Paper II that the HST/ACS
imaging used to derive the axis ratio in distant galaxies allows us to
reach the optical radius ($\sim 3.2 \times R_{d}$), which limits such
a bias. We note that extinction corrections are relatively small in
the K-band: in the distant sample, we find a mean/median correction of
0.11$\pm$0.05 mag ($1-\sigma$ dispersion), while in the local sample,
the mean (median) correction is found to be 0.15(0.14)$\pm$0.05, which
are consistent within the respective uncertainties.

\subsubsection{Effects related to the comparison to the local relation}
One important aspect about the evolution of the TFR is to control
systematic effects between the distant and local relations. These
considerations have led us to adopt as a local reference the TFR
derived by \cite{hammer07} for a complete subsample of the SDSS
\citep{pizagno07} (see Sect. 2.2). However, there are several
differences between the local and distant samples that could lead to
possible biases when comparing the corresponding TFRs.

First, kinematical data are obtained using different techniques (3D
spectroscopy vs. long-slit spectroscopy). We mitigated this effect by
restricting the local sample to galaxies having a well-defined RC (see
details in \citealt{hammer07}). This also mitigates the assumption
that both local and distant galaxies have RC shapes than can be well
described by an $\arctan$ (see Sect. 2.2 and 3.2). Second, a different
emission line is used for deriving the rotation velocity ([OII] and
H$_\alpha$). However, both emission lines are found to provide similar
estimates of the rotation velocity (e.g., \citealt{saintonge07}).
Third, in the local sample, $V_{80}$ (defined as the velocity at a
radius that encompasses 80\% of the light) is used as a proxy for
$V_{flat}$, while we directly used the asymptotic $V_0$ from the
$\arctan$ RC in the distant sample. $V_{80}$ is a reliable proxy for
$V_{flat}$, as shown in \cite{hammer07} (see also
\citealt{pizagno07}). Finally, one possible caveat about the use of
2MASS K-band magnitudes was recently pointed out by
\cite{noordermeer07}, i.e. that 2MASS underestimates the K-band
luminosity in relatively low surface brightness galaxies. We note that
this bias would go in the opposite trend compared to the evolution
seen in the TFRs, i.e., to shift the local relations towards brighter
magnitudes or larger stellar masses. However, we have also mitigated
this effect by restricting the local relation to galaxies having
$Log(V_{flat}) \geq 2.2$, i.e. to the most massive galaxies (see Sect.
2.2).

\subsubsection{Kinematic systematic uncertainty}
We restricted the analysis of the TFR evolution to well-relaxed RDs.
In distant TFR studies, it is often assumed that the (gas) rotation
velocity can be simply derived as the half-amplitude of the VF divided
by $\sin{(i)}$ (modulo spatial-resolution effects): an important
assumption behind this simple prescription is that the azimuthal
velocity component of the 3D velocity vector within the galaxy (i.e.,
the rotation) dominates its projection along the line-of-sight. This
assumption should always be checked \emph{a posteriori}, by showing
that residuals between a rotating-disk model and the observed VF are
small compared to the amplitude of rotation motions. In the case of
GIRAFFE RDs, there are few doubts that this assumption is correct,
because their modeled VFs match the observed ones (see Paper I).

Conversely, CK galaxies are clearly out of dynamical equilibrium (see
\citealt{flores06} and Paper I). As a consequence, one does no longer
know which component of the 3D velocity vector dominates its
projection along the line-of-sight. If these galaxies are associated
with mergers, as we will discuss below, deriving the rotation velocity
in this way is probably uncertain, if not meaningless. This is the
reason why in this study, we restricted the analysis of the TFR to
well-relaxed RDs.

In Sect. 3, we tested extensively the method of correction used to
correct $\Delta V_{obs}$ for spatial-resolution effects. All galaxies
were assumed to be RDs, regardless of their kinematical class (see
\citealt{puech06,puech07}). However, this provides uniform $\alpha$
values that are independent of kinematical class: a misclassification
of a galaxy has no impact on the way $\alpha$ is derived. This
approach helps to compare with long-slit spectroscopy results, where
all galaxies are implicitly assumed to be RDs (e.g.,
\citealt{conselice05}). As explained above, the fit to the TFR is
restricted to RDs only, and therefore this cannot affect the evolution
of the zero point of the TFR.

One might wonder whether or not the slight bias found in Fig.
\ref{compv} between early- and late-type galaxies could influence
significantly the results. Once translated into $\log{(V_{flat})}$
(using Eq. \ref{eqmk}), the offset of the distant TFR is found to be
-0.1 dex between $z\sim0.6$ and $z=0$, which is much larger than that
found in Fig. \ref{compv} for early-type galaxies ($\sim$0.025 dex).
Late-type galaxies have the opposite trend, which provides a -0.014
dex offset in Fig. \ref{compv}. It is impossible to explain all of the
shift in the distant TFR zero point in terms of such a morphological
bias. Looking at the TFR residuals against $B/T$ for RDs analyzed in
Paper II, we find no special trend, which excludes the presence of a
bias.

Another source of systematic uncertainty could be the limited spatial
coverage of the GIRAFFE IFU. Fig. \ref{compv} (see the black squares)
shows that when the plateau of the RC is not sampled by the IFU, the
recovered rotation velocity is underestimated on average by $\sim$0.03
dex. We note that this effect generally leads to the underestimation
of $V_{flat}$, which would increase the shift of the zero point
between the local and the distant TFRs. However, most of the RDs are
well spatially covered by the IFU (see \citealt{flores06} and Paper
I), and this could affect a few galaxies in our sample. It is thus
unlikely that such an effect could affect significantly the results in
a systematic way.

Finally, the most important possible systematic uncertainty likely
comes from our limited knowledge of the seeing during observations.
Individual variations from galaxy to galaxy leads to relatively small
random uncertainties (see Sect. 3.2). However, we assumed a uniform
value of 0.8 arcsec during the rotation-velocity correction process.
According to Fig. \ref{Figseeing}, changing the seeing in the
simulations from 0.8 to 1.0 arcsec implies a systematic effect of
-0.02 dex on the rotation velocity correction.

\subsubsection{Total systematic uncertainty}
The only possible systematic effect we can identify so far is that
associated with the kinematics, due to the correction for the rotation
velocity. Using Eq. \ref{eqmk}, this translates into a possible
systematic uncertainty of $\pm$0.14 mag in the 0.66 mag evolution of
the K-band TFR zero point between $z\sim0.6$ and $z=0$, which
represents 20\% of the shift.

\section{Discussion}

\subsection{Comparison with \cite{flores06}}
In \cite{flores06} (see also \citealt{puech07}), we used
hydro-dynamical simulations of an Sbc galaxy to infer a mean
correction factor $\alpha=1.2\pm0.05$ on $V_{max}$ for galaxies of
typical diameters between 2 and 3 arcsec: this value is less than that
found here using Monte-Carlo simulations. At first sight, it is
surprising to find a larger mean $\alpha$ for $V_{flat}$ than for
$V_{max}$ (since $V_{flat}\leq V_{max}$). The explanation is likely
linked to the fact that in \cite{flores06}, we did not consider all
possible ranges of size, PA, inclination, rotation velocity and RC
gradient to derive this mean correction, because we were limited by
the hydro-dynamical models of spiral galaxies available to us at that
time. Another difference is the exponential dependence of $\alpha$
with galaxy size found in \cite{flores06}, which is not reproduced
here. The reason for this is that in \cite{flores06}, we used the same
simulation of an Sbc galaxy to simulate distant galaxies of different
sizes, which induced an intrinsic correlation between the RC gradient
and the galaxy size. Our tests show that if we introduce this
correlation in the Monte-Carlo simulations, we then recover the
exponential variation of $\alpha$ with $R_{half}$.

\begin{figure}[h]
\centering \includegraphics[width=8cm]{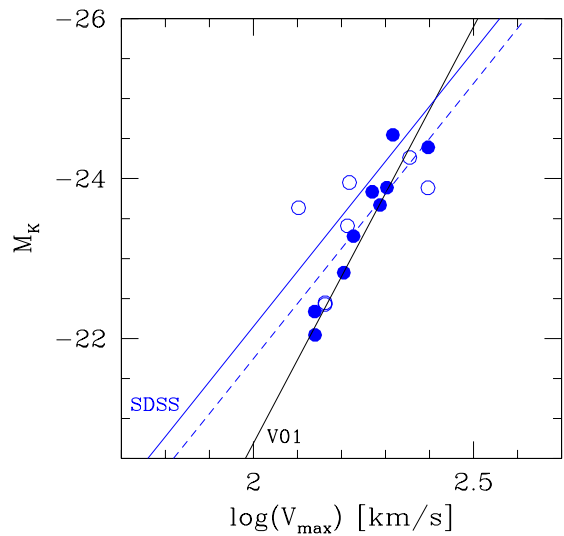}
\caption{K-band TFR derived following the \cite{flores06} methodology,
  for the RD subsample. The black line is the local relation of
  \cite{ver01}, i.e., the one used as a reference by \cite{flores06}.
  Note that K-band magnitudes in the distant sample have been
  converted into the Vega system using $M_K(Vega)=M_K(AB)-1.85$. For
  simplicity, we assume that the K' magnitudes of \cite{ver01} are
  roughly equivalent to those derived in the distant sample using the
  ISAAC Ks filter. The blue line is the SDSS local TFR used as a
  reference in this study (converted into the Vega system). The blue
  dash line is a linear fit to the distant RDs, which has a zero point
  0.4 mag lower than the local one, fixing the slope to the local
  value. Open symbols represent galaxies from the new CDFS sample,
  while full symbols represent galaxies used in \cite{flores06}.}
\label{tf_old}
\end{figure}

The brightening found in the K-band TFR was not initially detected by
\cite{flores06}. In Fig. \ref{tf_old}, we show the K-band TFR obtained
following the \cite{flores06} method to correct $\Delta V_{obs}$,
i.e., applying a constant correction factor of 1.2. The black line
shows the \cite{ver01} K-band local TFR using $V_{flat}$ as a
kinematical tracer for the rotation velocity, i.e., the relation used
as a local reference by \cite{flores06}. Both the local and distant
relation are then in good agreement: if we fix the slope of the
distant relation to the local one, we find a shift $\sim$ 0.1 mag
between the zero point of the distant and the local relations.
Therefore, we retrieve the \cite{flores06} results that no significant
evolution in zero point can be detected. The reason for such a
difference is twofold.

First, as detailed in \cite{hammer07}, we found a discrepancy between
the K-band TFR derived in the \cite{ver01} sample, compared to those
derived in two other local samples \citep{courteau97,pizagno07}. We
show in \cite{hammer07} that this discrepancy is due to the larger
fraction of faint and slow rotating galaxies in the \cite{ver01}
sample compared to the two other ones. Such galaxies have a larger gas
fraction, which results in a different slope at the low-mass end of
the TFR \citep{McGaugh05}. \cite{noordermeer07} derived a new K-band
local TFR: they found a slope similar to the one inferred by
\cite{hammer07} after restricting the \cite{ver01} TFR to the
high-mass end. In \cite{flores06}, the relation used as a local
reference was however the \cite{ver01} relation. If we refit the
distant RD subsample using the rotation velocities derived following
\cite{flores06}, but using the slope of the SDSS local relation (i.e.,
the one used in Fig. \ref{tf}, see the blue lines in Fig.
\ref{tf_old}), we find a shift $\sim$0.4 mag between the distant and
the local zero points. Compared to \cite{flores06}, we use a different
local sample, which provides a better control of systematic effects
that can occur in the comparison with the distant sample (see Sect.
4.2.2): this rigorous approach allows us to explain $\sim$ 60\% of the
TFR zero-point shift, previously hidden by spurious effects.

\begin{figure}[h]
\centering \includegraphics[width=8cm]{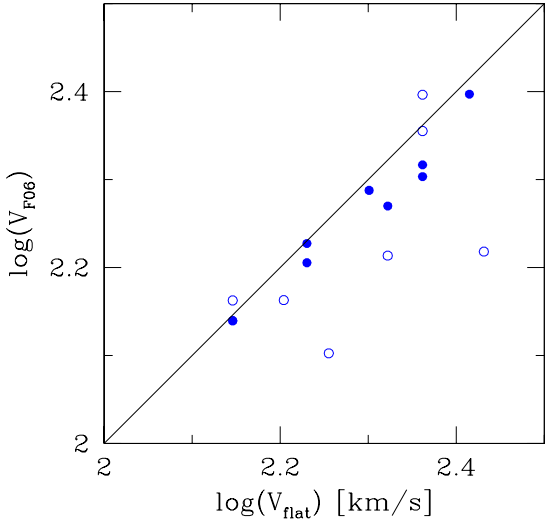}
\caption{Comparison between the rotation velocities $V_{F06}$
  obtained using the method of \cite{flores06} (i.e., a constant
  correction factor of 1.2) vs. rotation velocities $V_{flat}$
  derived using the new method used in this paper, for the subsample
  of RDs. $V_{F06}$ underestimates $V_{flat}$ by $\sim$11\% on
  average. Open symbols represent galaxies from the new CDFS sample,
  while full symbols represent galaxies used in \cite{flores06}.}
\label{vflatmax}
\end{figure}

Second, we have significantly improved the method for correcting the
rotation velocity since the preliminary work of \cite{flores06}. This
is illustrated in Fig. \ref{vflatmax}, where we compare the rotation
velocities derived following \cite{flores06} and those derived in this
study: this figure reveals that \cite{flores06} underestimated the
rotation velocity in RDs by 11\% on average, which corresponds to a
$\sim$0.05 dex shift in $\log{(V_{flat})}$, or $\sim$0.3 mag once
converted into a $M_K$ shift using Eq. \ref{eqmk}. Therefore, we
attribute the remaining $\sim$40\% of shift in the TFR zero point,
previously undetected, to the improvement in the rotation-velocity
derivation.

Finally, Fig. \ref{tf_old} provides us with a useful cross-check of
the new method used in this study to correct the rotation velocity. In
Fig. \ref{tf_old}, one RD galaxy (J033212.39-274353.6) appears to be
shifted to lower velocities compared to other distant rotating disks.
This galaxy was identified as having a particularly high $\alpha$
value in Sect. 3.3, while its data are well described by the TFR in
Fig. \ref{tf}. This suggests that the new method of correction appears
to be better suited for deriving rotation velocities: comparing Fig.
\ref{tf_old} with Fig. \ref{tf}, we note that the dispersion of the RD
subsample is significantly reduced using the new method of correction.

\subsection{Origin of the scatter in the distant relation}
Using the spatially-resolved kinematics provided by 3D spectroscopy,
we confirm earlier results that the dispersion of the distant
relations can be explained by the presence of dynamically non-relaxed
galaxies \citep{flores06}. These galaxies are found to be out of
dynamical equilibrium, with random motions instead of ordered motions,
especially for the most compact galaxies \citep{puech06, weiner06}.
Ongoing comparisons between hydrodynamical simulations of galaxy
mergers with GIRAFFE data reveal that they could be associated with
major mergers (\citealt{puech07}; Puech et al., in prep.). Such events
provide a natural link between CK galaxies and
morphologically-peculiar galaxies (see Paper II), which causes an
increase in the dispersion of the TFR at high redshift
\citep{kassin07,atkinson07}.

On the other hand, it is still unclear whether or not more quiescent
processes, such as secular gas accretion through internal processes
(bars), or cold-gas flows \citep{dekel06,keres05}, could explain such
anomalous kinematics on a relatively large spatial scales, especially
for CK galaxies (see Paper I). More work is required to compare
dynamical predictions of such events with observations.

\subsection{Comparison with the evolution of the K-band luminosity density}
Finding a 0.66 mag \emph{brightening} of the K-band TFR between
$z\sim0.6$ and $z=0$ appears to be quite surprising, given the
opposite trend seen in the evolution of the K-band luminosity density
in ``blue'' galaxies over the same redshift range. One may reasonably
assume that ``blue'' star-forming galaxies and emission-line galaxies
belong to the same population \citep{hammer97}: \cite{arnouts07} find
that the K-band luminosity density of blue star-forming galaxies
\emph{fades} by 0.5-0.6 mag from $z\sim0.6$ to $z=0$, which is exactly
the opposite trend. What appears to be a clear contradiction simply
reflects two different methodologies.

As we have detailed above, the TFR allows us to compare two
physically-connected populations of galaxies, i.e., the local spirals
and the distant rotating disks: the evolution of this relation
directly reflects the evolution of the K-band luminosity in
\emph{rotating disks of similar total mass} between $z\sim0.6$ and
$z=0$, assuming that the rotation velocity can be used as a proxy for
the total mass \citep{hammer07}. On the other hand, the color
selection used to select ``blue'' galaxies at $z=0$ and at $z\sim0.6$
can produce heterogeneous galaxy populations, as noticeable when
comparing their luminosity densities. Optical colors are well known to
be seriously affected by instantaneous star formation and extinction.
The evolution of the K-band luminosity density in ``blue'' galaxies
reflects the number evolution of ``blue'' galaxies, which are much
more numerous at high redshift (e.g., \citealt{ellis97}).

To check that there is no contradiction between the evolution of the
TFR zero point and the evolution of the K-band luminosity density, we
derived the K-band luminosity density in the complete GIRAFFE sample,
i.e., including all dynamical classes. As stated in Sect. 2.1, the
GIRAFFE sample is representative of $z\sim0.6$ emission-line galaxies
with $M_J(AB) \leq -20.3$ (i.e., with $M_{stellar} \geq
1.5\times10^{10}M_{\sun}$, see Paper I). We therefore expect a K-band
luminosity density in this sample that represents the luminosity
density of ``blue'' galaxies at these redshifts. We estimated the
K-band luminosity in the GIRAFFE sample to be $\log{(\rho _K)}=8.74$
L$_\odot$.Mpc$^{-3}$, which agree well with the K-band luminosity
density of ``blue'' star-forming galaxies found by \cite{arnouts07}.
This comparison confirms that the GIRAFFE sample is representative of
$z\sim0.6$ emission-line galaxies: the K-band luminosity density in
the GIRAFFE sample is consistent with the results inferred from
studies using far larger samples.

\subsection{Did distant rotating disks double their stellar-mass over
  the last 6 Gyr?} 

Once restricted to well-relaxed RDs, we find, between $z\sim0.6$ and
$z=0$, a shift in the TFR zero point of 0.66$\pm$0.14 mag in $M_K$, or
-0.1$\pm$0.02 dex in $\log{(V_{flat})}$. We consider the
interpretation of this shift in terms of galaxy evolution.

Could there be any so-called ``progenitor bias'' between distant RDs
and local relaxed spirals, which could imply that the latter would not
be the descendants of the former? Local RDs are found to be twice as
numerous as distant ones (see Paper II), which implies that some were
not in a relaxed dynamical state at $z\sim 0.6$ (being then PR or CK).
However, local intermediate-mass spirals have only a low probability
to have undergone a major merger since $z\sim 0.6$ (15-30\%, as
discussed in \citealt{puech07}): this means that most distant RDs must
dynamically evolve smoothly towards local relaxed spirals, which
implies that distant RDs are the progenitors of a majority of local
spirals. Therefore, it makes sense to interpret the evolution of the
RD-restricted TFR as an evolution in luminosity, rotation velocity, or
a combination of the two. We reiterate that we assume no evolution in
slope (see Sect. 4.1): a larger sample would be required to tackle
directly the possible evolution in slope of the TFR and determine the
details of the evolutionary path between distant and local RDs.

We discuss the possibility that this shift could correspond to a pure
luminosity brightening of 0.66$\pm$0.14 mag in RDs over the past 6
Gyr. Such a luminosity-brightening in RDs would correspond to a growth
in stellar-mass, which can be estimated in the following way. Between
$z\sim0.6$ and $z=0$, the evolution in $\log{(M_{stellar}/L_K)}$ is
found to range between 0.13 and 0.16 dex \citep{drory04,arnouts07},
depending on the selection criteria. Using
$\log{(M_{stellar})}=\log{(M_{stellar}/L_K)}+\log{(L_K)}$, one finally
finds a stellar-mass evolution of 0.39-0.42 dex. A more exhaustive
derivation, using the stellar-mass TFR, leads to a similar conclusion,
with an evolution in zero point of 0.36$^{0.21}_{-0.06}$ dex between
$z\sim0.6$ and $z=0$ (see Appendix A). On the other hand, the
evolution of the cosmic stellar-mass density over the same redshift
range is found to be $\sim$0.1-0.16 dex
\citep{dickinson03,drory04,arnouts07}. \cite{hammer05} found that most
of the present-day stellar-mass formed since $z=1$ occur in
intermediate-mass galaxies (see also \citealt{bell05}), which include
60\% of emission-line galaxies at $z\sim0.6$ \citep{hammer97}. Hence,
if we assume that, on average, quiescent galaxies do not evolve in
stellar-mass, one can estimate the growth in stellar-mass between
$z\sim0.6$ and $z=0$ in intermediate-mass, emission-line galaxies to
be $\sim$0.32-0.38 dex. Therefore, this interpretation of the TFR
zero-point shift as a pure luminosity evolution is roughly in
agreement with the evolution of the cosmic stellar-mass density over
the same redshift range. This interpretation is supported by the fact
that distant RDs are found to be LIRGs or at least star-bursting,
i.e., they are actively forming stars at very high rates (see Paper
II).

Interpreted in this way, the evolution of the zero point of the K-band
TFR reflects the growth in stellar-mass of the most active population
over the past 6 Gyr, i.e. star-forming intermediate-mass galaxies, by
a factor $\sim$2.5-2.6. This compares well with the gaseous O/H phase
abundance of $z\sim0.6$ emission-line galaxies, which is, on average,
half that of present-day spirals for a similar range of stellar mass
\citep{liang06}. Such an evolution in stellar mass would imply that
RDs converted an important amount of gas into stars over the past
6 Gyr. In other words, the main evolutionary path for RDs during the
past 6 Gyr would be conversion of gas into star through gas supply,
which is further supported by their relatively low gas-disk $V/\sigma$
values \citep{puech07} and the inside-out build-up of their stellar
disks (see a detailed discussion in Paper II).

The opposite interpretation is that the shift in the zero point of the
K-band TFR corresponds to a pure rotation-velocity evolution of
-0.1$\pm$0.02 dex between $z\sim0.6$ and $z=0$. Even in this case, the
observed evolution in $M_{stellar}/L_K$ over this redshift range still
implies, on average, a substantial growth in stellar mass of 0.13-0.16
dex in intermediate-mass galaxies (see above). As a consequence, the
stellar mass growth in \emph{quiescent} intermediate-mass galaxies
should be approximately similar because the stellar mass growth in the
entire population (i.e., quiescent or not) is observed to be similar
in this redshift range. This is clearly not what is observed, because
$\sim$80\% of the stellar mass formed since z=1 occurred in
star-forming galaxies (LIRGs, see \citealt{hammer05}). We can
therefore rule out a pure evolution along the velocity axis.

The last possibility is a combined evolution along both axes, i.e. a
simultaneous brightening in luminosity with a decrease in rotation
velocity $V_{rot}$. According to the Virial theorem, $V_{rot}^2$
scales as the ratio of the total mass enclosed within the optical
radius $R_{opt}$, over $R_{opt}$ ($\sim3.2R_d$ in both distant and
local galaxies, see Paper II). A decrease in $V_{rot}$ between
$z\sim0.6$ and $z=0$ would then imply a decrease in this ratio, which
in turn would imply that $R_{opt}$ (or $R_d$) increases faster than
the total mass over the same redshift range. Between $z\sim0.6$ and
$z=0$, $R_d$ does not seem to evolve strongly, at least in the RD
subsample (see Paper II; see also \citealt{puech07}). Therefore, only
a moderate increase in the total mass enclosed within $R_{opt}$ could
occur over this redshift range. Such a scenario would agree with
observations: as stated above, local intermediate-mass spirals have a
low probability to have undergone a major merger since $z\sim0.6$
(15-30\%, see \citealt{puech07} and references therein). Therefore,
most distant RDs should be already mostly assembled at $z\sim0.6$, and
should not undergo strong evolution in terms of total-mass from
$z\sim0.6$ to $z=0$.

To explore the real evolution along both axes, we compare our results
with the model of spherical gas accretion of \cite{birnboim07}.
\cite{birnboim07} present a model of accretion for a star-bursting
galaxy at $\sim$0.7 and describe its subsequent evolution in terms of
mass, down to $z=0$. Although it is not clear whether or not such a
model could be representative of all properties of distant RDs, we
assume that it can be used to constrain the average mass evolution in
a typical RD halo. In that case, the results of \cite{birnboim07}
would suggest that the baryonic mass in the disk remains constant
between $z\sim0.6$ and $z=0$, while the \emph{Virial} baryonic mass
roughly doubles (see their Fig. 2). On the other hand, \cite{conroy07}
found that the Virial-to-stellar mass ratio in intermediate-mass
galaxies is roughly constant between $z=1$ and $z=0$. Assuming that
the subsample of RDs follows the same trend, one can combine the
\cite{conroy07} observational results with the model of
\cite{birnboim07}, and find that RDs would roughly double their
stellar-mass between $z\sim0.6$ and $z=0$. Accounting for the
evolution in mass-to-light ratio over this redshift range, this would
translate into an evolution of 0.35-0.43 mag in luminosity, to be
compared with the 0.66 mag found in the evolution of the TFR zero
point. Once translated into $\log{(V_{flat})}$, this would allow a
-0.04 dex evolution along the velocity axis, between $z\sim0.6$ and
$z=0$.

In conclusion, we find that the most likely interpretation of the
evolution of the TFR zero point is the one in which this shift
reflects mostly a luminosity evolution of RDs. We estimated an upper
limit to the contribution of an evolution along the velocity axis to
be at most one half of the total shift of zero point. Such a
brightening of distant RDs over the past 6 Gyr would indicate a
doubling of their stellar-mass, which is independently supported by
their other dynamical and morphological properties.

Finally, this could suggest that the \emph{baryonic} (stars
\emph{plus} gas) TFR would not evolve with redshift: if one accounts
for the (average) two times larger gas fraction of $z\sim0.6$ galaxies
compared to $z=0$ \citep{liang06}, one finds that distant galaxies
roughly fall back onto the local smTFR (see Appendix A). Studies of
the local baryonic TFR have shown that galaxies having
$V_{flat}\leq$100 km/s systematically fall below the TFR defined by
more massive galaxies (e.g., \citealt{McGaugh05}). However, once the
gas fraction is accounted for, all galaxies follow the same
\emph{baryonic} TFR. Interpreted that way, our results suggest an
evolution of this threshold toward higher masses (velocities) at
higher redshifts, of at least $V_{flat}\sim$300 km/s at $z\sim0.6$.
This supports the idea that the baryonic TFR would be much more
``fundamental'' that the stellar-mass TFR \citep{McGaugh05}.

\section{Conclusion}
We have studied the evolution of the K-band TFR, using a
representative sample of 65 emission line, intermediate-mass galaxies
at $z\sim0.6$, unaffected by field-to-field variations within Poisson
statistics, and observed using 3D spectroscopy. We have presented and
tested a new method allowing us to safely recover $V_{flat}$ of
distant galaxies. We have also investigated possible sources of
systematic effects, which can be particularly important when studying
the evolution of the TFR. We have paid particular attention to the
analysis of both the local and distant samples in as similar as
possible ways. We draw the following conclusions:

\begin{description}
\item (1) The larger dispersion of the distant TFR is caused by
  galaxies with anomalous kinematics, ranging from perturbed rotators
  to very dynamically-disturbed galaxies. We find a positive and
  strong correlation between the complexity of the kinematics and the
  scatter contributed to the TF. Once restricted to relaxed rotating
  disks, the scatter of the TFR appears to not have evolved, which
  might suggest no evolution in slope;

\item (2) We detect for the first time a significant evolution of the
  K-band TFR zero point, which we attribute to an average brightening
  of $z\sim0.6$ galaxies by 0.66$\pm$0.14 mag. We attribute this
  evolution to the fact that selected distant galaxies are either
  starbursts or LIRGs. The distant emission-line rotating disks
  represent roughly one fourth to one fifth of present-day spiral
  progenitors, and one half of the whole population of $z\sim0.6$
  intermediate-mass rotating disks (see Paper II). Therefore, a
  significant part of spiral progenitors are doubling their stellar
  masses during the last $\sim$6 Gyr, in good agreement with former
  studies on the evolution of the mass-metallicity relation
  \citep{liang06}, which would suggest no evolution of the baryonic
  TFR with redshift;

\item (3) Current studies of the evolution of the TFR, even using
  spatially-resolved kinematics, are limited by important systematic
  uncertainties, which can be attributed to the limited spatial
  resolution of the kinematics, and to the derivation of the
  stellar-mass. These systematic uncertainties represent 20\% of the
  evolution in zero point of the K-band TFR.
\end{description}

Further progress in the study of the evolution of the TFR will benefit
from a forthcoming larger sample of RDs. This will allow us to reduce
the random uncertainties, allow the slope to vary during the fit, and
start studying TF residuals. Although we have limited the local sample
to galaxies having a well-defined rotation curve, a decisive answer
about the evolution of the TFR will require both local and distant
representative samples processed in the same way, from the
observational strategy (3D spectroscopy) to the data analysis itself,
as we have done in this paper. A similar analysis of a representative
local sample observed \emph{using 3D spectroscopy} is required to
confirm the above results. This point will be addressed in a future
study, using the local kinematic GHASP survey, whose analysis is
currently ongoing \citep{Epinat08}. Finally, an important limitation
remains linked to systematic uncertainties, which are mostly due to a
lack of spatial resolution. The only way to overcome this limitation
will be to develop a new generation of optical integral-field
spectrographs with higher spatial-resolution, i.e., assisted by
adaptive optics. In this respect, the future extremely large
telescopes will allow us to make a decisive leap forward in our
understanding of the TFR \citep{puech08}.

\appendix

\section{The evolution of the stellar-mass TFR}

\subsection{The local stellar-mass TFR}
Stellar masses $M_{stellar}$ were estimated from $M_{stellar}/L_K$
ratios using the method of \cite{bell03}. We used a solar luminosity
in the Ks-band of 3.28 (Vega) and assumed a ``diet'' Salpeter IMF
\citep{bell03}. This method takes advantage of the tight correlation
found between rest-frame optical colors and $M_{stellar}/L_K$ ratios,
assuming a universal IMF. These correlations are found to be
relatively insensitive to the details of galaxy SFH, dust content, and
metallicity \citep{bell01,bell03}, which implies that they are
invaluable for deriving stellar mass without being too sensitive to
the details of the stellar population synthesis models. We note that
following this method, $M_{stellar}/L_K$ ratios are corrected for the
amount of light due to red-giant stars using g-r colors.

According to \cite{bell03}, using this method, the total random
uncertainty on $\log{(M_{stellar}/M_\odot)}$ at $z\sim0$ should be lower
than 0.1 dex, and the systematic uncertainties due to galaxy ages,
dust, or bursts of star-formation can reach 0.1 dex. Finally, the
influence of TP-AGB stars in the derivation of stellar masses could
result in an overestimation of the stellar mass by $\sim$0.14 dex
\citep{maraston06,pozzetti07} in a systematic way.

Using this method, we converted the local K-band TFR in the subsample
of the SDSS (see Sect. 2.2) into stellar masses and found:
\begin{equation}
\log{(M_{stellar}/M_\odot)}=4.46\pm0.53+(2.8\pm0.23)\times \log{(V_{flat})},
\label{eqms}
\end{equation}
with $\sigma _{res}$=0.15 dex.

\subsection{The distant stellar-mass TFR}
To estimate stellar masses in the distant sample (see Table
\ref{ech}), we used the same method as in the local sample, i.e., the
K-band luminosity and B-V rest-frame colors. One important issue is
whether or not the same correlations between color and
$M_{stellar}/L_K$ ratios found at z=0 by \cite{bell03} can be directly
applied at higher redshift. \cite{borch06} showed that this appears
possible at least up to z$\sim$0.6 (see their Fig. 4), with an
associated random uncertainty of $\sim$0.3 dex, and an average
systematic uncertainty of up to -0.2 dex (i.e., towards an
overestimation of the stellar-mass at high $z$).

\begin{figure}
\centering
\includegraphics[width=8.9cm]{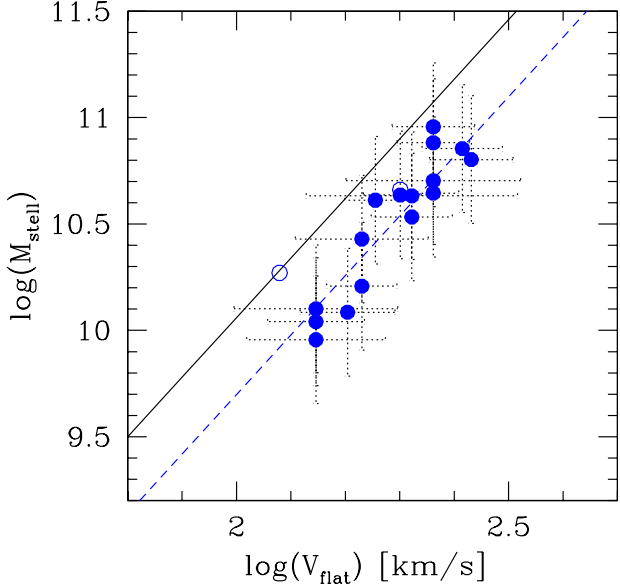}
\caption{Evolution of the stellar-mass TFR in the RD subsample (the
  two RD+ galaxies are represented with open blue dots). The black
  line is the local smTFR, while the blue dash-line represent a linear
  fit to the $z\sim0.6$ smTFR}
\label{smtf}
\end{figure}

We show the derived distant smTFR in Fig. \ref{smtf}. The smTFR shows
the same trend with the kinematic classification as the K-band TFR.
Restricting ourselves to RDs, and holding the slope constant, the
distant and local relations have similar dispersions ($\sigma
_{res}$=0.12 dex). If we allow the slope to vary during the fit, we
find a residual dispersion of 0.12 dex, suggesting again that assuming
no evolution in slope is acceptable. Maintaining the slope at its
local value, the distant smTFR zero point is found to be
4.10$\pm$0.16(random)$^{+0}_{-0.2}$(systematic, see above), i.e., 0.36
dex smaller than the local zero point (see Table \ref{fittfm}). A
Welcher's t-test gives a probability $\ll$1\% that the two relations
have the same zero point.

\begin{table*}
\centering
\caption{Fits to the local and distant smTFRs, using
  $\log{(M_{stellar}/M_\odot)}=a+b\times\log{(V_{flat})}$.}
\begin{tabular}{lccl}\hline\hline
K-band TFR        & Slope a        & Zero point b   & Comments         \\\hline
Local relation    & 2.8$\pm$0.23   & 4.46$\pm$0.53  & SDSS subsample  \\\hline
Distant RDs       & 2.8            & 4.10$\pm$0.16  & Using local slope\\
Distant RDs       & 3.28$\pm$1.10  & 3.01$\pm$2.47  & Slope free       \\\hline
Distant RD/RD+    & 2.8            & 4.13$\pm$0.15  & Using local slope\\
Distant RD/RD+    & 2.61$\pm$1.33  & 4.56$\pm$3.03  & Slope free       \\\hline\hline
\end{tabular}
\label{fittfm}
\end{table*}

\subsection{Stellar-mass systematic uncertainties}
The most important systematic uncertainty is that associated with the
mass-to-light ratios [M/L] predicted by stellar population synthesis
[SPS] models. Absolute values of M/L depend mostly on the SPS model
and IMF used, and more particularly on the prescriptions for the
TP-AGB stellar-evolution phase \citep{maraston05,maraston06}.
\cite{pozzetti07} compared M/L predictions between \cite{bruzual03}
and \cite{maraston05} SPS models and found a systematic difference of
-0.14 dex due to different prescriptions for TP-AGB stars. However,
they found that this systematic is constant with redshift, at least up
to $z\sim1.2$. In other words, at a given IMF, relative predictions
between two different redshifts of SPS models are much more robust in
terms of $M/L$ predictions.

\begin{figure}
\centering
\includegraphics[width=8.9cm]{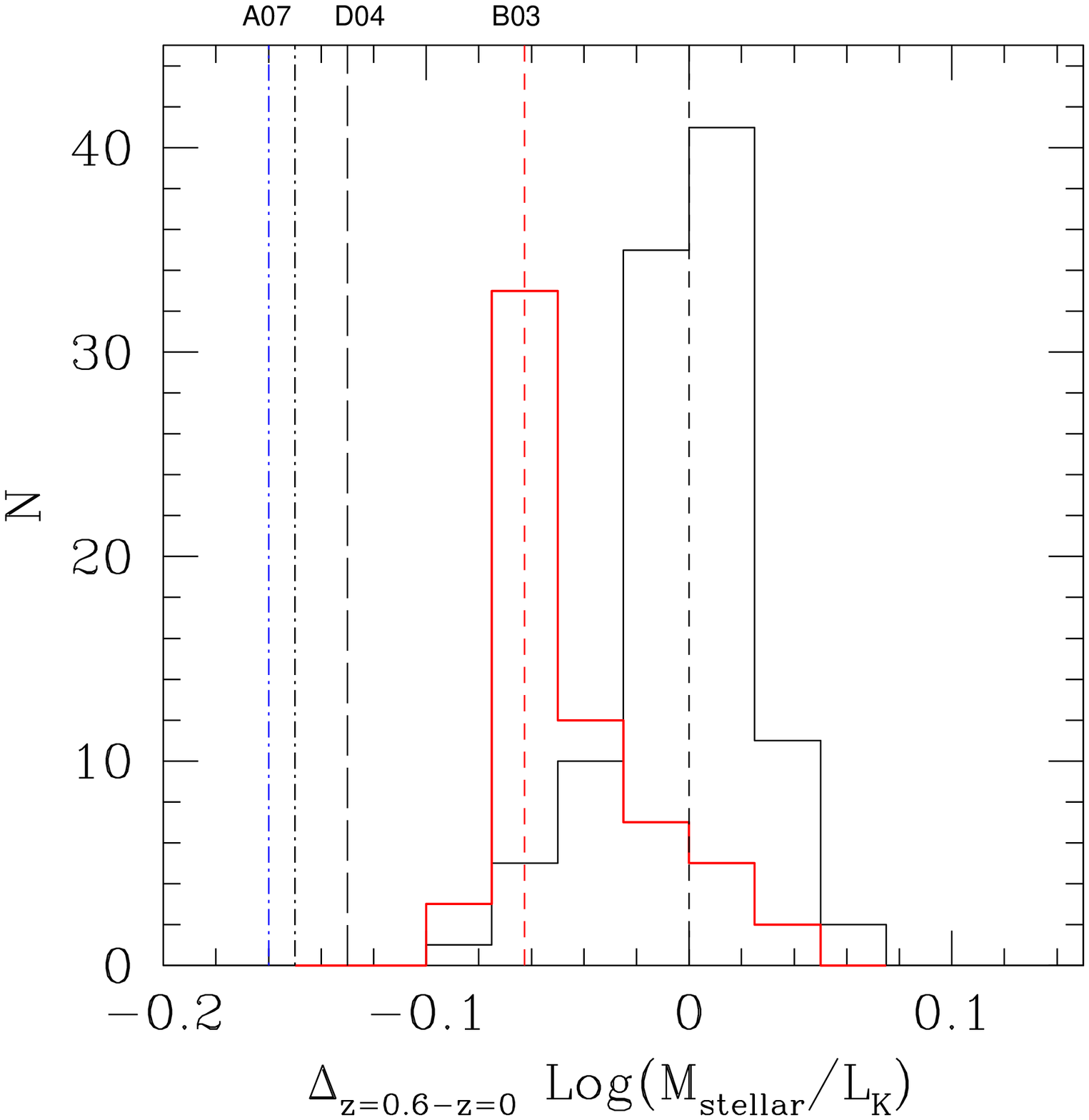}
\caption{Histograms of $\log{(M_{stellar}/L_K)}$ found in the local and
  distant samples of galaxies using the method of \cite{bell03}. Both
  histograms have been re-centered using the median found in the local
  samples, which allows us to directly infer the evolution of
  $\log{(M_{stellar}/L_K)}$ between $z\sim0.6$ and $z=0$, i.e., $\sim$0.625
  dex. Also shown, the evolution of $\log{(M_{stellar}/L_K)}$ found by
  \cite{drory04} in a sample of intermediate-mass galaxies (black
  long-dashed line), and \cite{arnouts07} in blue star-forming
  galaxies (blue mixed-line) or independently of the color (black
  mixed-line). }
\label{ml}
\end{figure}

We therefore focus in Fig. \ref{ml} on the evolution of
$\log{(M_{stellar}/L_K)}$ between $z\sim0.6$ and $z=0$: we show the
histograms of $\log{(M_{stellar}/L_K)}$ derived using the
\cite{bell03} method, both in the distant and local samples. Both
histograms were centered using the median $\log{(M_{stellar}/L_K)}$ of
the local sample: we find that $\log{(M_{stellar}/L_K)}$ evolves from
$z\sim0.6$ to $z=0$ by $\sim$0.06 dex. In this figure, we show other
determinations of the evolution of $\log{(M_{stellar}/L_K)}$ from the
literature. \cite{drory04}, using \cite{maraston98} SPS models, found
an evolution of 0.13 dex in galaxies with stellar masses between
4$\times10^{10}M_\odot$ and $10^{11}M_\odot$, i.e., in a range of
stellar-mass similar to GIRAFFE galaxies. This is similar to the 0.15
dex evolution found by \cite{arnouts07} in a flux-limited sample over
the same redshift range, using \cite{bruzual03} SPS models.
Furthermore, \cite{arnouts07} find an evolution of 0.16 dex once
restricted to a sample of blue star-forming galaxies. Because such
blue galaxies and emission-line galaxies probably belong to the same
populations \citep{hammer97}, we conclude that we might be
underestimating the evolution of $\log{(M_{stellar}/L_K)}$ by up to
$\sim$0.16-0.06=0.1 dex in a systematic way.

What is the origin of this systematic effect? As
$\log{(M_{stellar}/L_K)}$ depends mostly on color and not on mass,
\cite{bell03} used SPS models to predict their ``average''
correlations given a reasonable range of SPS parameters (e.g.,
metallicity, star formation histories). However, \cite{bell03} did not
explicitly fit the age but assumed instead a reference age of 12 Gyr
at $z=0$. This introduces a systematic difference with other results
where age is fitted explicitly. For instance, \cite{drory04} find an
average age of 8.1 Gyr at $z=0$. If one compares the median
$\log{(M_{stellar}/L_K)}$ found in the local SDSS sample used in this
study, with the average value found by \cite{drory04} in a similar
mass range, one finds a difference of -0.08 dex (which accounts for
the different IMF used). Similarly, we find a difference of -0.16 dex
between the median $\log{(M_{stellar}/L_K)}$ found in the GIRAFFE
sample, and the average value found by \cite{drory04} at $z=0.6$.
These values are consistent with the systematic error bars quoted
above. They suggest that we underestimate the evolution of
$\log{(M_{stellar}/L_K)}$ between $z=0$ and $z\sim0.6$ by -0.08 dex,
which is consistent with Fig. \ref{ml}.

In summary, these comparisons show that we are probably
underestimating the evolution of $\log{(M_{stellar}/L_K)}$ between $z=0$
and $z\sim0.6$ by -0.08 dex, and maybe up to -0.1 dex due to the use of
the simplified prescriptions for $M/L$ of \cite{bell03}. This directly
translates into a systematic effect of the evolution of smTFR zero
point by up to +0.1 dex.

\subsection{Total systematic uncertainty}
We summarize all possible sources of systematic uncertainty identified
so far in Table \ref{systsm}. We expressed all in terms of their
influence on the shift in the zero point between the distant and the
local smTFRs. Taking into account all these uncertainties, we find a
shift in the smTFR zero point of 0.36$^{0.21}_{-0.06}$ dex between
$z\sim0.6$ and $z=0$. Systematic uncertainties represent up to
$\sim$60\% (in dex) of the evolution detected in zero point, i.e.,
much larger than systematic uncertainties affecting the evolution of
the K-band TFR zero point. The derivation of stellar mass is highly
model-dependent, which implies a large additional uncertainty.
However, we note that these effects tend to overestimate the stellar
mass at high redshift, i.e., to minimize the evolution of the smTFR.
Using $\log{(M_{stellar}/L_K)}=\log{(M_{stellar})}-\log{(L_K)}$, one
finds that $\log{(M_{stellar}/L_K)}$ evolves by $\sim$0.1 dex
($=0.36-0.66/2.5$) between $z\sim0.6$ and $z=0$ in the subsample of
RDs. Therefore, the evolution in the zero point of the smTFR is quite
robust, in the sense that we might have minimized the evolution of
$M_{stellar}/L_K$ through the choice of method used to derive the
stellar mass (see Sect. A.3).

\begin{table*}
\centering
\caption{Identified systematic uncertainties that could impact the
  shift of zero point between the local and the distant smTFRs.
  Systematic uncertainties on $V_{flat}$ have been converted into
  $M_{stellar}$ using Eq. \ref{eqms}. Negative values tend to reduce
  the shift of zero point, while positive values have the opposite
  trend.}
\begin{tabular}{lcl}\hline\hline
Description & Bias on $M_{stellar}$ (dex) & Comments\\\hline
Photometry                  & $\sim$0      & Comparison EIS vs. ACS in the CDFS\\
M/L$_K$ Evolution with z    & +0.1         & Following Sect. A.3\\
IMF Evolution with z        & +0.05        & Following \cite{vandokkum07}\\
Velocity correction         & $\pm$0.06    & See Sect. 4.2.3\\
Total                       & -0.06/+0.21  & \\\hline\hline
\end{tabular}
\label{systsm}
\end{table*}

\begin{acknowledgements}
We thank all the GIRAFFE team at the Observatories of Paris and
Geneva, and ESO for this unique instrument.
\end{acknowledgements}

\end{document}